\documentclass[preprint,eqsecnum,aps]{revtex4}
\usepackage{graphicx}
\begin{document}


\title { Four-body cluster structure of $A=7-10$ double-$\Lambda$ 
hypernuclei}

\author{E. Hiyama}

\address{Institute of Particle and Nuclear Studies,
High Energy Accelerator Research Organization (KEK),
Tsukuba 305-0801, Japan} 

\author{M. Kamimura}

\address{Department of Physics, Kyushu University, 
Fukuoka 812-8581,Japan}
 
\author{T. Motoba}

\address{Physics Department, Brookhaven National Laboratory,
Upton, New York 11973\\
and Laboratory of Physics, Osaka Electro-Comm. University,
Neyagawa 572-8530, Japan\footnote{permanent address.}}

\author{T. Yamada}

\address{Laboratory of Physics, Kanto Gakuin University, 
Yokohama 236-8501, Japan}

\author{Y. Yamamoto}

\address{Physics Section, Tsuru University, Tsuru, Yamanashi 402-8555, 
Japan}
%
%
\begin{abstract}

Energy levels of the double-$\Lambda$ hypernuclei
$_\Lambda^{ }$$_\Lambda^7$He, 
$_\Lambda^{ }$$_\Lambda^7$Li, 
$_\Lambda^{ }$$_\Lambda^8$Li, 
$_\Lambda^{ }$$_\Lambda^9$Li, 
$_\Lambda^{ }$$_\Lambda^9$Be and 
$_\Lambda^{ }$$_\Lambda^{\!\!\!10}$Be
are predicted 
on the basis of the $\alpha+x+\Lambda
+\Lambda$ four-body model
with $x=n, p, d, t, ^3$He and $\alpha$, respectively.
Interactions between the constituent particles
are determined so as to reproduce reasonably
the  observed low-energy properties of the $\alpha +x$ nuclei 
$(^5{\rm He}, ~^5{\rm Li}, ~^6{\rm Li}, ~^7{\rm Li},
~^7{\rm Be}, ~^8{\rm Be})$
and  the existing data of $\Lambda$-binding energies of the
$x+\Lambda$  and $\alpha+x+\Lambda$ systems
$(^3_{\Lambda}{\rm H},
~^4_{\Lambda}{\rm H}, ~^5_{\Lambda}{\rm He},
~^6_{\Lambda}{\rm He}, ~^6_{\Lambda}{\rm Li},  ~^7_{\Lambda}{\rm Li},
~^8_{\Lambda}{\rm Li},  ~^8_{\Lambda}{\rm Be},
~^9_{\Lambda}{\rm Be})$. Here,
an effective $\Lambda \Lambda$ interaction is 
constructed 
so as to  to reproduce,
within the $\alpha+\Lambda+\Lambda$ model, the
$B_{\Lambda \Lambda}$ of 
$_\Lambda^{ }$$_\Lambda^6$He which was discovered recently in 
the NAGARA event of the emulsion experiment.
With no  adjustable parameters for the 
$\alpha+x+\Lambda+\Lambda$ systems,
$B_{\Lambda \Lambda}$ of the ground  
and bound excited states of 
the double-$\Lambda$ hypernuclei with $A=7 - 10$
are accurately calculated with the Gaussian-basis 
coupled-rearrangement-channel method.
The {\it Demachi-Yanagi} event, observed recently 
for $_\Lambda^{ }$$_\Lambda^{\!\!\!10}$Be, is interpreted as
observation of its $2^+$ excited state
on the basis of the present calculation.
Structure change of the $\alpha+x$ core nuclei 
due to the participation of the
$\Lambda$ particles is found to be substantially large and
it plays an important role 
in estimating the $\Lambda \Lambda$ bond energies of those
hypernuclei.

\end{abstract}


\maketitle


\section{Introduction}
A recent finding of the double-$\Lambda$ hypernucleus
$_\Lambda^{ }$$_\Lambda^6$He, which is called as NAGARA event in
the KEK-E373 experiment~\cite{Nagara}, has a great impact not only
on the study of baryon-baryon interactions in the strangeness
$S=-2$ sector but also on the study of dynamics of many-body systems
with multi-strangeness. The importance of this event is
attributed to the well-defined explanation of the process and
the high quality experimental value of the $\Lambda \Lambda$ binding
energy $B_{\Lambda \Lambda}=7.25 \pm 0.19\pm ^{0.18}_{0.11}$ MeV 
~\cite{Nagara},
which leads to a smaller $\Lambda \Lambda$ binding,
$\Delta B_{\Lambda \Lambda}=1.01 \pm 0.20 \pm ^{0.18}_{0.11}$ MeV,
than the previous understanding.
 Sometimes the emulsion events include
ambiguities related with serious difficulty of identifying emission
of neutral particles such as neutrons and $\gamma$-rays. In the
NAGARA event, however, the production of 
$_\Lambda^{ }$$_\Lambda^6$He
has been uniquely identified free from such an ambiguity on the basis
of the observation of sequential weak decays.

Historically, in the 1960's, there 
appeared two reports on the observation of
double-$\Lambda$ hypernuclei, 
$_\Lambda^{ }$$_\Lambda^{\!\!\!10}$Be~\cite{Danysz63}
and $_\Lambda^{ }$$_\Lambda^6$He~\cite{Prowse66}, but the reality
of the latter case was considered doubtful~\cite{Dalitz89}.
Two decades later the modern emulsion-counter hybrid technique has been
applied in the KEK-E176 experiment~\cite{Aoki91}, 
in which a new double-$\Lambda$
hypernucleus event was found but no unique identification was given
so far: One explanation as $_\Lambda^{ }$$_\Lambda^{\!\!\!10}$Be 
leads to a repulsive $\Lambda \Lambda$ interaction 
($\Delta B_{\Lambda \Lambda}<0$), 
while the other possibility
involving $_\Lambda^{ }$$_\Lambda^{\!\!\!13}$B leads to an attractive
$\Lambda \Lambda$ interaction ~\cite{Dover91,Yamamoto91}. 
If the latter is the case, 
the extracted strength of the $\Lambda \Lambda$ interaction is
attractive with $\Delta B_{\Lambda \Lambda} \simeq 4$ MeV.
Although the latter option seems consistent with
the old data of $_\Lambda^{ }$$_\Lambda^{\!\!\!10}$Be~\cite{Danysz63},
the substantially attractive 
$\Lambda \Lambda$ interaction has not been convincing.

In the strangeness nuclear physics, the most fundamental
problem is to recognize various facets of interactions among
octet baryons ($N$, $\Lambda$, $\Sigma$, $\Xi$) in a unified way.
Our detailed knowledge for the $S=0$ $NN$ sector is
based on the rich data of $NN$ scatterings as well as nuclear phenomena.
Recent studies for $S=-1$ many-body systems such as $\Lambda$
hypernuclei have clarified interesting features of $\Lambda N$
and $\Sigma N$ interactions in spite of scarce data of the free-space
scatterings.
On the other hand, for 
baryon-baryon interactions with $S=-2$ sectors, concerned presently,
experimental information has been highly limited due to the extreme
difficulties  of two-body scattering experiments.
Therefore the observed $\Lambda \Lambda$ bond energies of
double-$\Lambda$ hypernuclei should be the 
 most reliable source for the  $S=-2$ interaction, and
such data play a decisive role in
determining the strength of underlying $\Lambda \Lambda$ interactions.

In view of this relevance and the experimental
situation,
the NAGARA event is an epoch-making one which
provides us with a new and firm basis for understanding
the double-$\Lambda$ hypernuclei.
In recent years several experiments to produce
$S=-2$ systems (E176 and E373 at KEK, E885 and E906 at BNL)
have been performed and some of the data analyses
are still in progress to get novel information
on the $S=-2$ interactions.

In this exciting situation of the experimental study,
it is needed to perform careful theoretical calculations of
double-$\Lambda$ hypernuclei with refreshed viewpoints.
As one of the motivations of the present work,
we think it necessary and timely to put the
NAGARA data of $_\Lambda^{ }$$_\Lambda^6$He binding energy
as a new standard basis for a systematic study of a series of
several double-$\Lambda$ species.
Secondly, in order to extract information on the $\Lambda \Lambda$
interactions precisely, here we emphasize that
hypernuclear calculations should be complete and
realistic enough to leave structural ambiguity as negligibly
as possible.
All the dynamical changes due to successive $\Lambda$ participation
should be also taken faithfully.
To meet these requirement we explore light $p$-shell double-$\Lambda$
hypernuclei ($A=6-10)$ comprehensively using the microscopic
three- and four-body models.
Thirdly, by these systematic and realistic calculations,
we will give reliable prediction of not
only the ground-state binding energies but also
possible excited-state energies,
which encourages double-$\Lambda$ hypernuclear 
spectroscopic study in near future.

So far several cluster models have appeared to estimate
the ground-state binding energies of double-$\Lambda$ species:
Based on the old data of $_\Lambda^{ }$$_\Lambda^6$He and
$_\Lambda^{ }$$_\Lambda^{\!\!\!10}$Be, 
Takaki {\it et al.}~\cite{Takaki89} applied
a simplified version of the $\alpha+x+ \Lambda + \Lambda$
cluster model to $A=6 - 10$ systems in which
they put several angular momentum restriction
and neglected rearrangement channels.
Bodmer {\it et al.}~\cite{Bodmer84,Bodmer87} 
performed variational Monte Carlo
calculations for $\alpha+ \Lambda+\Lambda$ 
and $\alpha+\alpha+ \Lambda+ \Lambda$ to
investigate consistency between the $\Lambda \Lambda$-binding energies,
$B_{\Lambda \Lambda}$($_\Lambda^{ }$$_\Lambda^6$He) and
$B_{\Lambda \Lambda}$($_\Lambda^{ }$$_\Lambda^{\!\!\!10}$Be),
although their old data should be now updated.
In the latest stage of this work, we
encountered with the Faddeev-Yakubovsky calculations
of $_\Lambda^{ }$$_\Lambda^6$He 
and $_\Lambda^{ }$$_\Lambda^{\!\!\!10}$Be
by Filikhin and Gal~\cite{Gal01} who restricted
the equations within the $s$-wave.
They compared the results  with our previous
cluster-model calculation~\cite{Hiyama97} 
which was performed with wider model
space but the stronger $\Lambda \Lambda$ interaction
strength. 
In our previous work~\cite{Hiyama97}, 
$\Lambda \Lambda$ binding energies
have been  calculated for $_\Lambda^{ }$$_\Lambda^6$He 
and $_\Lambda^{ }$$_\Lambda^{\!\!\!10}$Be 
in the framework of the
$\alpha+\Lambda+\Lambda$ three-body model and the
$\alpha+\alpha+ \Lambda+ \Lambda$ four-body model, respectively,
where the adopted $\Lambda \Lambda$ interaction is taken
to be considerably attractive on the basis of the
traditional interpretation for the double-$\Lambda$ events.

In the present work, by noting the importance of
the NAGARA data, we extend this four-body model to
more general cases consisting of $\alpha+x+ \Lambda+ \Lambda$ 
systems with $x=n,p,d,t,^3$He and $\alpha$ 
($_\Lambda^{ }$$_\Lambda^7$He, $_\Lambda^{ }$$_\Lambda^7$Li,
$_\Lambda^{ }$$_\Lambda^8$Li, $_\Lambda^{ }$$_\Lambda^9$Li,
$_\Lambda^{ }$$_\Lambda^9$Be and
$_\Lambda^{ }$$_\Lambda^{\!\!\!10}$Be), where 
nuclear core parts 
are quite well represented  by $\alpha+x$ cluster models
(for example in Ref.~\cite{Furutani80}). 
 Here we remark that the extensive calculations are presented
for the first time for $A=7-9$ double-$\Lambda$ species and that 
the old predictions for $_\Lambda^{ }$$_\Lambda^6$He
and $_\Lambda^{ }$$_\Lambda^{\!\!\!10}$Be have been
updated in a unified way.
The four-body calculations are accurately
performed by using the Jacobian-coordinate Gaussian-basis
method of Refs.
\cite{Kami88,Kame89,Hiyama95,Hiya96,Hiyama00,Hiyama02,Kamada01}
with all the rearrangement channels taken into account.
In our model, structure changes
of nuclear cores caused by added one and two $\Lambda$ particles
are treated precisely. Namely, we take into account
the rearrangement effects on $\Lambda \Lambda$
bond energies induced by changes of nuclear cores.
It is worthwhile to point out that the important effects of
core-excitations and core-rearrangement are lacking
in the frozen-core approximation used often for
calculations of double-$\Lambda$ hypernuclei.

In our model, it is possible to determine the 
$\alpha x$ and $\Lambda x$ interactions
so as to reproduce all the existing
binding energies of subsystems
($\alpha+x$, $x+\Lambda$, $\alpha+x+ \Lambda$
and $\alpha+\Lambda+\Lambda$)
in an $\alpha+x+\Lambda+\Lambda$ system,
where that of $\alpha+\Lambda+\Lambda$ is given by the NAGARA event.
This feature  is important to discuss 
the energy levels of the double-$\Lambda$
hypernuclei and to extract
the $\Lambda \Lambda$ interactions because the ambiguities
of $NN$ and $\Lambda N$ effective interactions are renormalized
by fitting the known binding energies of subsystems phenomenologically.
Our analysis is performed systematically for ground and
bound excited states of the series of $\alpha+x+\Lambda+\Lambda$ systems
with no more adjustable parameters in this stage,
so that these predictions offer
an important guidance to interpret coming
double-$\Lambda$ events in the  experiments and then to
determine the level structure and
the $\Lambda \Lambda$ interaction unambiguously.

In Section II, the calculational method with microscopic 
$\alpha+x+ \Lambda + \Lambda$ four-body model
is described. In Section III, 
the  interactions are introduced.
Calculated results are
presented and discussed in Section IV. Summary is given in Section V.



%
\section{Model and method}

In Ref.~\cite{Hiyama97}, the present authors already studied
$_\Lambda^{ }$$_\Lambda^6$He and 
$_\Lambda^{ }$$_\Lambda^{\!\!\!10}$Be 
with the use of 
$\alpha+\Lambda+\Lambda$ three-body model and
$\alpha +\alpha+\Lambda+\Lambda$ 
four-body model, respectively.
In the same manner, we study in this work 
the double-$\Lambda$ hypernuclei 
$_\Lambda^{ }$$_\Lambda^7$He, 
$_\Lambda^{ }$$_\Lambda^7$Li, 
$_\Lambda^{ }$$_\Lambda^8$Li, 
$_\Lambda^{ }$$_\Lambda^9$Li, 
$_\Lambda^{ }$$_\Lambda^9$Be and 
$_\Lambda^{ }$$_\Lambda^{\!\!\!10}$Be 
on the basis of   
the $\alpha+x+\Lambda+\Lambda$ four-body model with
$x=n, p, d, t, ^3$He and $\alpha$, respectively. 
The $d, t (^3{\rm He})$ and $\alpha$ clusters are assumed  
to be inert having the
$(0s)^2$, $(0s)^3$ and $(0s)^4$ shell-model 
configurations, respectively, 
and are denoted by $\Phi_s(x)$ with spin $s\, (=1, \frac{1}{2}$
or 0, respectively). 

All the nine sets of the Jacobian coordinates of 
the four-body systems are illustrated
in Fig. 1 in which
we further take into account the antisymmetrization  
between two $\Lambda$ particles and
the symmetrization  
between two $\alpha$ clusters when $x=\alpha$.
The total Hamiltonian and the Schr\"{o}dinger equation 
are given by
\begin{eqnarray}
     H=T+\sum_{(a,b)}V_{a b}
      +V_{\rm Pauli} \; ,  \\
    ( H - E ) \, \Psi_{JM}(^A_{\Lambda \Lambda}{\rm Z})  = 0 \;,
\end{eqnarray}
where $T$ is the kinetic-energy operator and 
$V_{ab}$ is the interaction
between the constituent particle-pair $a$ and $b$.
The Pauli principle between the nucleons 
belonging to $\alpha$ and $x$ clusters
is taken into account by the 
Pauli projection operator 
$V_{\rm Pauli}$ which is explained  
in the next section as well as  $V_{a b}$. 
The total wave function is described
as a sum of amplitudes of the rearrangement channels 
$(c=1-9)$ of Fig. 1 in the LS coupling scheme:
\begin{eqnarray}
      \Psi_{JM}\!\!&(&\!\!^A_{\Lambda \Lambda}{\rm Z})
       =  \sum_{c=1}^{9} 
      \sum_{n,N,\nu}  \sum_{l,L,\lambda} 
       \sum_{S,\Sigma,I,K}
       C^{(c)}_{nlNL\nu\lambda S\Sigma IK} \nonumber  \\
      &  \times & {\cal A}_\Lambda{\cal S}_\alpha 
      \Bigg[  
       \Phi(\alpha) \left[\Phi_s(x)
    \; [ \chi_{\frac{1}{2}}(\Lambda_1) 
       \chi_{\frac{1}{2}}(\Lambda_2)
     ]_S \right]_\Sigma  \nonumber \\
  & \times  &    \left[ 
     [\: \phi^{(c)}_{nl}({\bf r}_c) 
         \psi^{(c)}_{NL}({\bf R}_c)
     ]_I \; \xi^{(c)}_{\nu\lambda} (\mbox{\boldmath $\rho$}_c) 
        \right]_{K}  \Bigg]_{JM}  \;  .
\end{eqnarray}

Here the operator  ${\cal A}_{\Lambda}$
stands for antisymmetrization between the two 
$\Lambda$ particles,
and ${\cal S}_{\alpha}$ is the symmetrization operator
for exchange between $\alpha$ clusters when $x=\alpha$. 
$\chi_{\frac{1}{2}}(\Lambda_i)$ is the spin function of the
$i$-th $\Lambda$ particle.
Following the Jacobian-coordinate 
coupled-rearrangement-channel Gaussian-basis variational
method of Refs.
\cite{Kami88,Kame89,Hiyama95,Hiya96,Hiyama00,Hiyama02,Kamada01},
we take the functional form of 
$\phi_{nlm}({\bf r})$, 
$\psi_{NLM}({\bf R})$ and
$\xi^{(c)}_{\nu\lambda\mu} ({\boldmath \rho}_c)$ as
\begin{eqnarray}
      \phi_{nlm}({\bf r})
      &=&
      r^l \, e^{-(r/r_n)^2} 
       Y_{lm}({\widehat {\bf r}})  \;  ,
 \nonumber \\
      \psi_{NLM}({\bf R})
      &=&
       R^L \, e^{-(R/R_N)^2} 
       Y_{LM}({\widehat {\bf R}})  \;  ,
 \nonumber \\
      \xi_{\nu\lambda\mu}(\mbox{\boldmath $\rho$})
      &=&
       \rho^\lambda \, e^{-(\rho/\rho_\nu)^2} 
       Y_{\lambda\mu}({\widehat {\mbox{\boldmath $\rho$}}})  \; ,
\end{eqnarray}
where the Gaussian range parameters are chosen to lie
in geometrical progressions:
\begin{eqnarray}
      r_n
      &=&
      r_1 a^{n-1} \qquad \enspace 
      (n=1 - n_{\rm max}) \; ,
\nonumber\\
      R_N
      &=&
      R_1 A^{N-1} \quad 
     (N \! =1 - N_{\rm max}) \; ,  
\nonumber\\
      \rho_\nu
      &=&
      \rho_1 \alpha^{\nu-1} \qquad 
     (\nu \! =1 - \nu_{\rm max}) \; .  
\end{eqnarray}
These basis functions have been verified to be suited for
describing both short-range correlations 
and long-range tail behaviour of few-body systems
\cite{Kami88,Kame89,Hiyama95,Hiya96,Hiyama00,Hiyama02,Kamada01}.
  The eigenenergy $E$  in Eq.(2.2) and the 
coefficients $C$ in Eq.(2.3) are to be determined by 
the Rayleigh-Litz variational method. 

For the angular-momentum space of the wave function,
the approximation with $l, L, \lambda \leq 2$ was found to be enough 
in getting satisfactory convergence of the binding energies
of the states concerned presently. 
Note that no truncation is taken of the {\it interactions}
in the  angular-momentum space.  As for the numbers
of the Gaussian basis,
$n_{\rm max}, N_{\rm max}$ and 
$ \nu_{\rm max}$, $4 -10$ are enough. 

  As far as the single $\Lambda$ hypernuclei 
$^6_{\Lambda}$He, $^6_{\Lambda}$Li, 
$^7_{\Lambda}$Li, 
$^8_{\Lambda}$Li, 
$^8_{\Lambda}$Be and 
$^9_{\Lambda}$Be 
are concerned,
the  wave functions  
are described by Eq.(2.3) but with one of $\Lambda$ particles omitted.
As for the core nucleus itself, $\alpha+x$, the wave function
is given by 
\begin{equation}
      \Psi_{JM}(\alpha+x) = \sum_{n,l}  C_{nl}  {\cal S}_\alpha 
   \Phi(\alpha) \big[\Phi_s(x)  \phi_{nl}({\bf r}) \big]_{JM} \; .
\end{equation}



\section{Interactions}

In the study of double-$\Lambda$ hypernuclei based on
the $\alpha + x + \Lambda + \Lambda$ four-body model, 
it is absolutely necessary and important
to examine, before the four-body calculation,
that the model with the interactions adopted
is able to reproduce reasonably well
the following observed quantities: 
(i) energies of the low-lying states
and scattering phase shifts of the $\alpha+x$ nuclear systems,
(ii) $B_\Lambda$ of hypernuclei composed of
$x+\Lambda$, $x$ being $d, t, ^3$He, $\alpha$,
(iii) $B_\Lambda$ of hypernuclei
composed of $\alpha+ x +\Lambda$, $x$ being 
$ n, p, d, t,^3$He, $\alpha$ and
(iv) $B_{\Lambda \Lambda}$ of  $_\Lambda^{ }$$_\Lambda^6$He
$=\alpha+\Lambda+\Lambda$.
We emphasize that this severe examination were successfully 
done in the present model as mentioned below.
This encourages  us to perform the four-body calculations,
with no adjustable parameters at this stage,
with high reliability of the results expected.

\subsection{Pauli principle between $\alpha$ and $x$ clusters}

The Pauli principle between nucleons belonging to 
$\alpha$ and $x$ clusters is taken into account by the 
orthogonality condition model
(OCM)\cite{ocm}.
The OCM projection operator
$V_{\rm Pauli}$ is represented 
by 
\begin{equation}
        V_{\rm Pauli}=\lim_{\lambda\to\infty}
        \lambda \, \sum_f
       |\phi_f({\bf r}_{\alpha x}) \rangle \langle
        \phi_f({\bf r}'_{\alpha x})| 
\end{equation}
which rules out the amplitude of the Pauli-forbidden 
$\alpha-x$ relative states $\phi_f({\bf r}_{\alpha x})$ 
 from the four-body 
total wave function \cite{Kukulin84}.  
The forbidden states are $f=0S$ for $x=n (p)$, 
$f=\{0S, 0P\}$ for $x=d$, 
 $f=\{0S,  1S, 0P, 0D\}$ for $x=t (^3$He)
and $f=\{0S,  1S, 0D\}$ for $x=\alpha$.
The Gaussian range parameter $b$ of the single-particle $0s$ orbit 
in the $\alpha$ particle is taken to be $b=1.358$ fm
so as to reproduce the size of the $\alpha$ particle. 
The same size is assumed for clusters $x=d, t $ and $ ^3$He
to manage the Pauli principle avoiding the calculational difficulty.
In the actual calculations, the strength $\lambda$
for $V_{\rm Pauli}$ is taken to be 
10$^5$ MeV which is large enough
to push up away the unphysical forbidden 
states in the very high energy
region with keeping the physical states unchanged. 
Usefulness of this Pauli operator method of OCM has been 
verified in many cluster-model calculations.

In some calculations
\cite{Bodmer84,Bodmer87,Gal01,Gal02,c12}
of three-body systems including
two or three $\alpha$ clusters, use is made of an
$\alpha \alpha$ potential with 
a strong repulsive core 
\cite{Bodmerxx}
so as to describe the Pauli exclusion role which
prevents the two $\alpha$ cluster from overlapping.  
But, it is well known 
\cite{Horiuchi} 
that this approximate 
prescription of the Pauli principle is not suited for
the case where the presence of the third particle makes the
two $\alpha$ clusters come closer to each other; 
in other words, the off-energy-shell
behaviour of the repulsive potential is not appropriate 
in the three-body system.  Moreover, there is no available potential 
reported for the $\alpha \,x$ systems ($x=n, p, d, t $ and $ ^3$He)
of this type.
Therefore, we do not employ this prescription in
the present systematic study of the structure change of 
the $\alpha \,x$ systems
due to the addition of $\Lambda$ particles. We take the
orthogonality condition model instead which is suited
even for the case of heavy overlapping between the two 
clusters. 


\subsection{$\alpha \,x$ interactions}

As for the potentials $V_{\alpha x}$ between the clusters
$\alpha$ and $x$, we employ those which have been
often used in the OCM-based cluster-model study of
light nuclei. Namely, they are
the $V_{\alpha N}$ potential introduced in Ref.
\cite{Kanada79}, 
the $V_{\alpha d}$ and $V_{\alpha t}$ 
potentials given in Ref.\cite{Furutani80}
and the $V_{\alpha \alpha}$  
potential used in Ref.
\cite{Hasegawa71} 
which  reproduce reasonably well the low-lying states
and low-energy scattering phase shifts of the $\alpha \, x$ 
systems. 
The potentials are described in the following 
parity-dependent form with the central and spin-orbit terms:
\begin{eqnarray}
&V&_{\!\!\!\!\!\alpha x}(r) =  \sum_{i=1}^{i_{\rm max}}
       V_i \: e^{-\beta_i r^2}
        + \sum_{i=1}^{i'_{\rm max}}
        (-)^lV_i^{\rm p} \: e^{-\beta_i^{\rm p} r^2} \nonumber \\  
         &  + & \big[\sum_{i=1}^{i''_{\rm max}}
           V_i^{ls} \: e^{-\gamma_i r^2}
           +  \sum_{i=1}^{i'''_{\rm max}}(-)^lV_i^{ls,{\rm p}}
        \: e^{-\gamma_i^{\rm p} r^2} \; \big] \;
        \mbox{\boldmath $l$}\cdot
               \mbox{\boldmath $s$}_x \;, \nonumber \\
\end{eqnarray}
where \mbox{\boldmath $l$} is the relative angular momentum 
between $\alpha$ and $x$,  and 
\mbox{\boldmath $s$}$_x$ 
is the spin of  $x$.  In the $\alpha \alpha$ system
the spin-orbit term is missing and the odd wave is forbidden
by the Pauli principle. 
The additional Coulomb potentials are constructed
by folding the $pp$ Coulomb force into the proton densities of
the $\alpha$ and $x$ clusters.
The parameters in Eq.(3.2) are listed in Table I (we slightly 
modified the strength of the central force in 
$V_{\alpha d}$ and that of the spin-orbit force in
$V_{\alpha t}$ to obtain better agreement with the 
energy levels of $^6$Li and $^7$Li, respectively).


\subsection{$\Lambda \,x$  interactions}

We derive the interaction between the $\Lambda$ particle and
the $x$ cluster by folding the G-matrix type $YN$ 
interaction (the YNG interaction) 
into the density of the $x$ cluster in the same manner
of our previous work on the double-$\Lambda$ hypernuclei
\cite{Hiyama97}.  
  The YNG interactions between $\Lambda$ and $N$
are derived from the $YN$ OBE models
as follows:
First the G-matrix equation is solved in nuclear matter
at each $k_{\rm F}$, where the 
so called QTQ prescription is adopted for simplicity.
Next the resulting G-matrix 
is simulated by a three-range Gaussian
form with the strengths as a function of $k_{\rm F}$. 
Obtained YNG interactions are given in 
Ref.\cite{Yama94} 
as
\begin{eqnarray}
              v_{\Lambda N}\!\!\!&(&\!\!\!\!r ; k_F)  =  \sum_{i=1}^{3}
             \Big[(v_{0,\rm even}^{(i)} + 
            v_{\sigma \sigma,\rm even}^{(i)}
              \mbox{\boldmath $\sigma$}_{\Lambda}\cdot
               \mbox{\boldmath $\sigma$}_N )
               \frac{1+P_r}{2}
                 \nonumber \\
             &+& (v_{0,\rm odd}^{(i)}
                + v_{\sigma \sigma,\rm odd}^{(i)}
                \mbox{\boldmath $\sigma$}_{\Lambda}\cdot
               \mbox{\boldmath $\sigma$}_N )
               \frac{1-P_r}{2} \Big] 
             \: e^{-\mu_i r^2}  , 
\end{eqnarray}
where $P_r$ is the space exchange (Majorana) operator.
The strengths $v^{(i)}$ 
are represented
as   quadratic functions of $k_{\rm F}$ ;
see Eq.(2.7) of  
Ref. \cite{Yama94} 
and Table V of
Ref. \cite{Hiyama97} 
for various original $YN$ interactions.
In the present work, we employ the Nijmegen model D interaction
(ND).

  The $\Lambda x$ interaction is derived 
by folding the above $v_{\Lambda N}(r; k_F)$ interaction 
into the $x$-cluster wave function.
The $k_{\rm F}$  depends on the mass number of the cluster $x$.
Because of the operator $P_r$ 
in Eq.(3.3), 
the resultant $\Lambda x$ potential becomes
nonlocal, the explicit form of  which is given
in Appendix of 
Ref.\cite{Hiyama97}. 
We summarize the functional form of the local and
nonlocal parts of the $\Lambda \,x$ potentials as
\begin{eqnarray}
      & V&_{\!\!\!\!\!\Lambda x}({\bf r},{\bf r'}) = 
            \sum_{i=1}^{3}
(V_i + V_i^{\rm s}  \mbox{\boldmath $s$}_{\Lambda}\cdot
              \mbox{\boldmath $s$}_x)             
 \: e^{-\beta_i r^2}
         \delta( {\bf r}-{\bf r'}) 
                 \nonumber \\
      &+& \sum_{i=1}^{3}
(U_i + U_i^{\rm s} \mbox{\boldmath $s$}_{\Lambda}\cdot
               \mbox{\boldmath $s$}_x )             
        \: e^{-\gamma_i({\bf r}+{\bf r'})^2
        -\delta_i({\bf r}-{\bf r'})^2}  \; ,
\end{eqnarray}
where  \mbox{\boldmath $s$}$_{\Lambda}=$
\mbox{\boldmath $\sigma$}$_{\Lambda}/2$.
      Table II lists the parameters in Eq.(3.4) for
(a)  $\Lambda \,\alpha$ interaction,
(b)  $\Lambda \, t \:(\Lambda ^3$He) interaction and
(c)  $\Lambda \,d$ interaction.
They were determined 
in the following manner:

i) {\it  $\Lambda \alpha$ interaction}:
The $\Lambda N$ spin-spin part
vanishes by the folding into the $\alpha$ particle.
The odd-force contribution is negligible to
the $\Lambda$-binding energy of $^5_\Lambda$He. 
We determined the  $k_{\rm F}$ parameter
as $k_{\rm F}=0.925\: {\rm fm}^{-1}$ in order 
to reproduce this binding energy (3.12 MeV) 
within the $\alpha+\Lambda$ two-body model.
The $\Lambda N$ odd-force having the same $k_{\rm F}$
was determined  by tuning the magnitude of $v_{0, \rm odd}^{(3)}$
so as to reproduce,
within the $\alpha+\alpha+\Lambda$ model,
the $\Lambda$-binding  energy of 
the $1/2^+$ ground state of $^9_\Lambda$Be.

ii) {\it  $\Lambda d$ interaction}:
We determined the value of $k_{\rm F}=0.84 \: {\rm fm}^{-1}$ 
by fitting the experimental $\Lambda$-binding energy of
the $1/2^+$ ground state of 
$^3_\Lambda$H within the $d+\Lambda$ model where 
the $\Lambda N$ odd force 
plays a negligible role.
The odd force was determined, with the same $k_{\rm F}$ kept,
by reproducing the $\Lambda$-binding 
energies of the $1/2_1^+$ and $3/2_1^+$ states 
of $^7_\Lambda$Li within the
$\alpha+d+\Lambda$ model; we  tuned 
$v_{0, \rm odd}^{(2)}$ and $v_{\sigma\sigma, \rm odd}^{(2)}$.

iii) {\it  $\Lambda t$ interaction}:
The experimental $\Lambda$-binding energies of 
the $0^+$ and $1^+$ states 
of $^4_\Lambda$H were used to 
determine the even force of the $\Lambda N$ interaction. 
The magnitude of 
$k_{\rm F}$ and $v_{\sigma\sigma,\; \rm even}^{(2)}$
were adjusted to reproduce the energies,
$k_{\rm F}$ being $0.84 \: {\rm fm}^{-1}$.
This value of $k_{\rm F}$ was substituted
into the $k_{\rm F}$ used in the odd force
of the $\Lambda N$ interaction of the $\Lambda d$ interaction
with no other  change. The resulting $\Lambda t$ interaction
reproduces, by chance, the $\Lambda$-binding energy of 
the $1^+$ ground state of $^8_\Lambda$Li
within the $\alpha+t+\Lambda$ model;
the calculated energy is 6.80 MeV while the
observed one is 6.80 $\pm$ 0.03 MeV.

\subsection{$\Lambda N$ interaction in 
$_\Lambda^{ }$$_\Lambda^7$He ($_\Lambda^{ }$$_\Lambda^7$Li) }

In the study of $_\Lambda^{ }$$_\Lambda^7$He
($_\Lambda^{ }$$_\Lambda^7$Li) with the
$\alpha+N+\Lambda+ \Lambda$ model and of the subsystem
$^6_\Lambda{\rm He}\:(^6_\Lambda{\rm Li})$
with the $\alpha+N+\Lambda$ model, it is inadequate to use
the G-matrix type $\Lambda N$ interaction because $\Lambda N$
correlations are fully taken into account in our model space.
Here, we employ a simple free-space $\Lambda N$ interaction
with a three-range Gaussian form, which simulates
the Nijmegen model F (NF) $\Lambda N$ interaction. Here, the
$\Sigma N$ channel coupling contribution is renormalized into
the $\Lambda N$ single channel with the closure approximation.
The even- and odd-state parts of our $\Lambda N$ interaction
are represented as follows:
\begin{eqnarray}
             &V\!\!\!\!&_{\Lambda N}(r)  =  \sum_{i=1}^{3}
             \Big[(v^{\rm even}_i + 
            v^{\rm even, \sigma}_i
              \mbox{\boldmath $\sigma$}_{\Lambda}\cdot
               \mbox{\boldmath $\sigma$}_N )
               \frac{1+P_r}{2}
                 \nonumber \\
             & &+\, (v^{\rm odd}_i
                + v^{\rm odd, \sigma}_i
                \mbox{\boldmath $\sigma$}_{\Lambda}\cdot
               \mbox{\boldmath $\sigma$}_N )
               \frac{1-P_r}{2} \Big] 
             \: e^{-\mu_i r^2} \; . 
\label{inta}
\end{eqnarray}
First, the parameters are dertemined so as to simulate the
$\Lambda N$ scattering phase shifts calculated with NF.
Next, the second-range strengths $v^{\rm even}_2$ and
$v^{\rm even, \sigma}_2$ are adjusted so as to reproduce the
$\Lambda$ binding energies of the $0^+$ and $1^+$ states of
$^4_\Lambda$H with the use of the $N+N+N+\Lambda$ four-body model.
Furthermore,
strengths $v^{\rm odd}_2$ and $v^{\rm odd, \sigma}_2$
are adjusted within the framework of
$\alpha+ \Lambda+ n+p$ four-body model
so as to reproduce the observed binding energies of
the ground-state spin doublet, $1/2^+$ and $3/2^+$ of
$^7_{\Lambda}$Li.
Our resulting parameters in Eq.(3.5) are listed in Table III.
We further found that the energy of the ground state of
$^6_\Lambda$He ($^6_\Lambda$Li) measured from the
$^5_\Lambda{\rm He}-N$ threshold can be well reproduced with
our $\Lambda N$ interaction
in the $\alpha+N+\Lambda$ three-body calculation;
for $^6_\Lambda$He ($^6_\Lambda$Li),
the calculated energy is $-0.17$ MeV (0.57 MeV), while the
observed one is $-0.17$ MeV (0.59 MeV).


\subsection{$\Lambda \, \Lambda$ interactions}

In the present model, since the $\Lambda \Lambda$ relative
motion is solved rigorously including the short-range
correlations, it is not adequate to use the $\Lambda \Lambda$
$G$-matrix interaction given in Ref. \cite{Yama94}.
However, our $\Lambda \Lambda$ interaction 
to be used in the present calculation should be still
considered as an effective interaction, since the
couplings to $\Xi N$ and $\Sigma \Sigma$ channels are
not treated explicitly.  Thus we employ the
$\Lambda \Lambda$ interaction  represented in
the following three-range Gaussian form:
\begin{equation}
    v_{\Lambda \Lambda}(r)=\sum_{i=1}^{3}
   (v_i + v_i^\sigma
\mbox{\boldmath $\sigma$}_{\Lambda}\cdot
    \mbox{\boldmath $\sigma$}_{\Lambda})
               \: e^{-\mu_i r^2} .
\end{equation}

It is enlightening here to keep some linkage to the
OBE models in determining the interaction parameters
$\mu_i, v_i$ and $v_i^\sigma (i=1-3)$.
In our previous work on $_\Lambda^{ }$$_\Lambda^6$He and 
$_\Lambda^{ }$$_\Lambda^{\!\!\!10}$Be
\cite{Hiyama97}, the interaction parameters
were chosen so as to simulate the $\Lambda \Lambda$ sector of
the ND interaction which is a reasonable model
for the strong attraction suggested by the old
interpretation for double-$\Lambda$ hypernuclei.
The characteristic feature of ND is that there is
only a scalar singlet instead of a scalar nonet, which gives
strongly attractive contribution in $\Lambda \Lambda$
as well as $NN$.

The other versions of the Nijmegen models with a scalar nonet
lead to much weaker $\Lambda \Lambda$ attractions, which seems to
be appropriate for the weak $\Lambda \Lambda$ binding indicated
by the NAGARA event. The NF is the simplest
among these versions, which is adopted here as a guidance to
construct our $\Lambda \Lambda$ interaction: The outer two
components  of the above Gaussian potential $(i=1,2)$ are
determined so as to simulate the $\Lambda \Lambda$ sector of NF,
and then the strength of the core part $(i=3)$ is adjusted so as
to reproduce the experimental value of
$B_{\Lambda \Lambda}(^{\ 6}_{\Lambda \Lambda}$He).
 The obtained values of parameters are given
in Table IV. It is interesting that the resulting
$\Lambda \Lambda$ interaction is almost equal to the interaction
obtained by multiplying a factor 0.5 on the above ND-simulating
interaction employed in Ref.\cite{Hiyama97}.


\section{Results and Discussions}

Let us show the calculated results for a series
of double-$\Lambda$ hypernuclei with $\alpha+x+ \Lambda +\Lambda$
structures ($x=0, n,p,d,t,^3{\rm He}, \alpha)$ 
studied in the microscopic four-body cluster model.
In order to understand the role of two $\Lambda$ particles attached to
the core nuclei, it is useful to compare the obtained level  structures
of the $\alpha+x+ \Lambda + \Lambda$ double-$\Lambda$ hypernuclei with
those of the $\alpha+x$ nuclei and the 
$\alpha+x+ \Lambda$ single-$\Lambda$ hypernuclei.
Then, we can see clearly how the ground and excited states of
$\alpha+x$ nuclei are changed due to the participation 
of $\Lambda$ particles.
It should be noted again here that, in the model description of
$\alpha+x+ \Lambda +\Lambda$, 
the observed low-energy properties of the
$\alpha+x$ nuclei and the existing 
$\Lambda$-binding energies of the $x+ \Lambda$ and 
$\alpha+x+ \Lambda$ hypernuclei
have been  reproduced accurately enough to give reliable
predictions for the  double-$\Lambda$ hypernuclei
with no adjustable parameters of the interactions
in the four-body calculations.
It offers the most reliable ground for our cluster
model that the threshold energies for every partition into
sub-cluster systems are assured to be reproduced.


\subsection{Energy spectra}
In Figs. $2 - 7$, the calculated 
level structure of $\alpha+x$ core nuclei, 
$\alpha+x+ \Lambda$  hypernuclei
and $\alpha+x+ \Lambda + \Lambda$ 
hypernuclei are illustrated side by side.
There are shown all the ground and bound excited states of
double-$\Lambda$ hypernuclei predicted in the present model.
In these figures, one sees clearly that
injection of one  and two $\Lambda$ particles 
leads to  stronger binding of the whole system
and prediction of more bound states.
But, there is no bound 
'$p$-orbit' of $\Lambda$ particle in 
single- and double-$\Lambda$ hypernuclei
with $A \leq 10$.
In the bound states of double-$\Lambda$ hypernuclei, 
two $\Lambda$ particles are coupled  to 
$S=0$ and therefore the spins and  parities are the same as 
those of its nuclear core.

Table V summarizes the calculated ground-state energies for
the double-$\Lambda$ hypernuclei including
the $2^+$ excited state of $_\Lambda^{ }$$_\Lambda^{\!\!\!10}$Be.
The results are expressed in terms of two quantities:
One is the total energy measured from the
breakup threshold of $\alpha +x+ \Lambda + \Lambda$
which is denoted as $E_{\Lambda \Lambda}$.
The other is 
$B_{\Lambda \Lambda}$ which is the 
binding energy of two $\Lambda$ particles 
with respect to 
the ground-state nuclear core $\alpha+x$.

The calculated values of $B_{\Lambda \Lambda}$ can be compared with
some experimental data, though they are quite
limited at present.
The most recent and clear
data of the NAGARA event is used as a basic  input
of our model so that our $\Lambda \Lambda$ interaction is
adjusted  to reproduce the experimental value  
$B_{\Lambda \Lambda}^{\rm exp}(^{\ 6}_{\Lambda\Lambda}$He)$
=7.25 \pm 0.19\pm ^{0.18}_{0.11}$ MeV~\cite{Nagara}. 
It is of particular interest to 
compare the present result with  another
data which is not used in the fitting procedure.
There is an event found in the E373 experiment, named as
{\it Demachi-Yanagi} event \cite{Ichi99,Ichi01};
the most probable interpretation of this event is a production
of  a bound state of 
$_\Lambda^{ }$$_\Lambda^{\!\!\!10}$Be
having $B^{\rm exp}_{\Lambda \Lambda}
=12.33 \pm^{0.35}_{0.21}$ MeV
which is obtained by assuming 
$B^{\rm exp}_{\Xi}=0.15 \pm ^{0.3}_{0.1}$ MeV.
In the emulsion analysis
there is no direct evidence for the production of
$^{10}_{\Lambda \Lambda}$Be in an excited state.
However, if the produced $^{10}_{\Lambda \Lambda}$Be 
is interpreted to be
in the ground state, the resultant 
$\Lambda \Lambda$ bond energy becomes repulsive, 
contradictorily to the NAGARA event. From 
the viewpoint of the present study,
the {\it Demachi-Yanagi} event can be interpreted most probably as
observation of the $2^+$ excited state in 
$^{10}_{\Lambda \Lambda}{\rm Be}$; 
our calculated value of 
$B_{\Lambda \Lambda}(^{10}_{\Lambda \Lambda}{\rm Be}(2^+)$ )
is 12.28 MeV, which agrees with the above experimental value.
This good agreement suggests that our level structures
calculated systematically are predictive and useful for coming
events expected to be found in the further analysis for the
E373 data.
Now it should be stressed that the above experimental data of
$_\Lambda^{ }$$_\Lambda^{\!\!\!10}$Be$(2^+)$ 
leads to no information for the ground-state value of
$B_{\Lambda \Lambda}$ unless the theoretical value
(2.86 MeV in our case) of the excitation energy of 
$_\Lambda^{ }$$_\Lambda^{\!\!\!10}$Be$(2^+)$ 
is utilized.

On the other hand, the old experiment
by Danysz {\it et al.} 
\cite{Danysz63} 
on the pionic decay of
$^{10}_{\Lambda \Lambda} {\rm Be} (0^+)$
$\rightarrow $ $^9_{\Lambda}{\rm Be}(1/2^+)+p+ \pi^-$
gave 
$B_{\Lambda \Lambda}^{\rm exp} (^{10}_{\Lambda \Lambda}{\rm Be} (0^+))
=17.7 \pm 0.4$ MeV.
This value has been used for a long time, which means the strongly
attractive $\Lambda \Lambda$ interaction.
However, it should be noted that the authors
also suggested the possibility of another decay 
$^{10}_{\Lambda \Lambda}{\rm Be} (0^+)
\rightarrow \;$ $^9_{\Lambda}{\rm Be}(3/2^+, 5/2^+)+p+ \pi^-$
(Table 5 of Ref.\cite{Danysz63}); 
the same was  pointed out in Ref. \cite{Gal02}, too.
In this case, the value of  
$B^{\rm exp}_{\Lambda \Lambda}$
($_\Lambda^{ }$$_\Lambda^{\!\!\!10}$Be $(0^+)$) 
is modified to 14.6 $\pm$ 0.4 MeV, which is obtained
by using the excitation energy of
$^9_{\Lambda}{\rm Be}(3/2^+, 5/2^+)$ to be 3.05 MeV \cite{Akikawa02}.
This modified value  turns out 
to be not contradictory with our calculated value, 15.14 MeV.
A similar re-interpretation, with the hypernuclear excited states
taken into account, 
may be  needed also for the E176 event
which was identified as $_\Lambda^{ }$$_\Lambda^{\!\!\!13}$B
($_\Lambda^{ }$$_\Lambda^{\!\!\!10}$Be) with the strongly 
attractive (repulsive)  ${\Lambda \Lambda}$
interaction.

Thus, we have understood the consistency between the
experimental data and our theoretical results of
$_\Lambda^{ }$$_\Lambda^{\!\!\!10}$Be.
We, therefore, discuss on the level structures of 
double-$\Lambda$ hypernuclei in more  detail.
As seen in  Figs. $2 - 7$ and Table V, 
the $\Lambda$ particle  plays a glue-like
role so that a whole system becomes of
stronger binding.
This effect in a double-$\Lambda$ nucleus is more enhanced than
that in the corresponding single-$\Lambda$ nucleus.
One can see a typical example in the case of 
$_\Lambda^{ }$$_\Lambda^7$Li in Fig. 3.
For the unbound nuclear system of $^5$Li, a single 
$\Lambda$ cannot make a bound system of
$^6_{\Lambda}$Li, but, the addition of one 
more $\Lambda$ particle
leads to a bound system of $_\Lambda^{ }$$_\Lambda^7$Li 
whose ground state is of
weakly-binding with respect to the 
 $_\Lambda^{ }$$_\Lambda^6$He$+p$ threshold.
 
The bound excited states of double-$\Lambda$ hypernuclei predicted
in the present cluster model are summarized as follows:
In $_\Lambda^{ }$$_\Lambda^7$He and $_\Lambda^{ }$$_\Lambda^7$Li,
the ground states are both bound but
no excited states are predicted.
It is needless to say, there are no bound excited
states in  double-$\Lambda$
hypernuclei with $A \leq 6$ since
there is no bound excited state in their core nuclei.
The lightest double-$\Lambda$ hypernucleus that has
at least one excited state is $_\Lambda^{ }$$_\Lambda^8$Li. 
In $_\Lambda^{ }$$_\Lambda^8$Li
we predict two $T=0$ excited
states in the bound-state region.
It is expected to have a $T=1, 0^+$ bound excited state in 
$_\Lambda^{ }$$_\Lambda^8$Li
which corresponds to the $T=1, 0^+$ state in $^6$Li nucleus  
at $E_{\rm x}=3.56 $ MeV, but the state 
is not shown in Fig. 4 because the $T=1$ state may have five-body
structure and is out of scope of
the present cluster model.
We predict three bound excited states in
$_\Lambda^{ }$$_\Lambda^9$Li ($_\Lambda^{ }$$_\Lambda^9$Be).
There is one bound excited state in $_\Lambda^{ }$$_\Lambda^{\!\!\!10}$Be
as mentioned before.
It will be challenging to discover these excited states one by one
as well as the ground states. 

\subsection{Dynamical change of the core nucleus}
It is interesting to look at the  dynamical change of 
the $\alpha+x$ nuclear cores 
which occurs due to the successive participation of
two $\Lambda$ particles.
The possibility that a nuclear core shrinks due to
an attached $\Lambda$ particle has been theoretically
pointed out using the $\alpha+x+ \Lambda$ cluster model
of light $p$-shell $\Lambda$ hypernuclei
~\cite{Motoba83}. 
An updated prediction 
\cite{Hiyama7Li} 
was made specifically on a shrinkage 
in size by 21\% to be seen in  $^7_{\Lambda}$Li.
The recent measurement of $\gamma$-ray transition
rate in $^7_{\Lambda}$Li~\cite{TamuraTanidaE419} 
has confirmed quantitatively the shrinkage effect 
predicted in both the old calculation and
the updated one.  
It is quite reasonable, therefore, that in 
a double-$\Lambda$ hypernucleus the participation
of one more  $\Lambda$ particle can induce  further 
shrinkage of the nuclear core. 
Such an effect has been also
investigated systematically using the molecular orbital model for
$^{8+n}_{n \Lambda}$Be $(n=1-4)=
\alpha + \alpha +n \Lambda$ \cite{Miyahara83}. 

In order to see such shrinkage effect, 
we show three physical quantities:
first in Table VI we list the 
r.m.s. distance between  $\alpha$ and $x$,
${\bar r}_{\alpha x} $.
As the number of the $\Lambda$ particles increases,
${\bar r}_{\alpha x} $ turns out to shrink significantly
due to the glue-like role of the attached  $\Lambda$ particles.
For example, one sees 
${\bar r}_{\alpha x}$ changes as 
$4.11 \rightarrow 3.44 \rightarrow 3.16$~fm
for $^6{\rm Li} \rightarrow ^7_{\Lambda}$Li$\rightarrow$ 
$_\Lambda^{ }$$_\Lambda^8$Li.
Participation of the second $\Lambda$ gives
rise to about 8 \% reduction of 
${\bar r}_{\alpha x}$ except $x=n$.
Secondly, in more detail, it is worthwhile to
demonstrate in Fig. 8 the change of the $\alpha-n$
two-body density (correlation function)
$\rho(r_{\alpha n}$) 
in the $^5$He, $^6_{\Lambda}$He and $_\Lambda^{ }$$_\Lambda^7$He
when $\Lambda$ particles
participate successively,
which again manifests the shrinkage effect.
Thirdly, 
this shrinkage effect is seen in
the large change of the expectation
value of the relative kinetic energy,  
$<T_{\alpha x}>$, and that of the potential energy, 
$<V_{\alpha x}>$, in the $\alpha- x$ subsystems.
When the $\alpha$ and $x$ clusters approach to each other,
the increase of $<T_{\alpha x}>$ overcomes the gain
of $<V_{\alpha x}>$,
and the sum $<T_{\alpha x}+V_{\alpha x}>$
increases appreciably.
In spite of this energy loss in
the $\alpha-x$ core system, 
the core shrinkage is realized by the stronger energy gain of
the $\Lambda - \alpha$ and $\Lambda - x$ parts.

\subsection{$\Lambda \Lambda$ bond energy}
In Fig. 9 we reveal the contributions of the $\Lambda \Lambda$
interaction to the total binding energies of double-$\Lambda$ 
hypernuclei $^A_{\Lambda \Lambda}$Z.
Here the calculated values of 
$B_{\Lambda \Lambda}(^A_{\Lambda \Lambda}$Z)
in the ground states are shown by closed circles.
In order to extract the contribution of
the $\Lambda \Lambda$ interaction, we perform the same
calculations by putting $V_{\Lambda \Lambda}=0$.
The obtained values are denoted as 
$B_{\Lambda \Lambda}(^A_{\Lambda \Lambda}$Z; $V_{\Lambda \Lambda}=0$)
and shown by open circles in the figure.
It should be noted that the effect of the dynamical change
of the $\alpha +x$ core 
due to the $\Lambda N$ interactions is included in
the four-body estimate of $B_{\Lambda \Lambda}$ 
and $B_{\Lambda \Lambda}$($V_{\Lambda \Lambda}=0$).
Since the $\Lambda \Lambda$ interaction
is not so strong compared with the $\Lambda N$ interaction,
the core-rearrangement effects included in
$B_{\Lambda \Lambda}$
and $B_{\Lambda \Lambda}$($V_{\Lambda \Lambda}=0$)
are similar to each other.
Then, naturally the pure effect of the $\Lambda \Lambda$
interaction is given by the difference
\begin{equation}
{\cal V}^{\rm bond}_{\Lambda \Lambda}
(^A_{\Lambda \Lambda}{\rm Z}) \equiv
B_{\Lambda \Lambda}(^A_{\Lambda \Lambda}{\rm Z})
-B_{\Lambda \Lambda}(^A_{\Lambda \Lambda}{\rm Z};
 V_{\Lambda \Lambda}=0).
\label{eq:DeltaBLL2}
\end{equation}
We consider 
${\cal V}^{\rm bond}_{\Lambda \Lambda}$
as the $\Lambda \Lambda$ bond energy 
which should be determined essentially 
by the strength of the $\Lambda \Lambda$ interaction.
Now in Fig. 9, we find that
the magnitude of ${\cal V}^{\rm bond}_{\Lambda \Lambda}$, 
the energy difference between the closed and open circles,
is almost constant at $\sim$ 1 MeV for all the 
double-$\Lambda$ hypernuclei with $A=6 - 10$.
The detailed  values of ${\cal V}^{\rm bond}_{\Lambda \Lambda}$
are listed in Table V.

So far the following intuitive formula has been often
used to estimate the $\Lambda \Lambda$ interaction strength:
\begin{equation}
\Delta B_{\Lambda \Lambda}(^A_{\Lambda \Lambda}{\rm Z}) \equiv
B_{\Lambda \Lambda}(^A_{\Lambda \Lambda}{\rm Z})
-2B_{\Lambda}(^{A-1}_{\Lambda}{\rm Z}).
\label{eq:DeltaBLL}
\end{equation}
It is worthwhile to point out the problems underlying in this
formula: This expression includes three problems 
which come from i) the mass-polarization term
of the three-body kinetic-energy operator,
ii) the $\Lambda N$ spin-spin interaction
and iii) the dynamical change of the core nuclear structure.

The problem i)  is stated as follows:
In the $\alpha + \Lambda + \Lambda$
three-body model for $_\Lambda^{ }$$_\Lambda^6$He
(generally, "$\alpha$"  may be replaced by 
"spinless frozen-core nucleus"),
if one takes the {\it non}-Jacobian coordinate set
${\bf r}_{\alpha\Lambda_1}$ and ${\bf r}_{\alpha\Lambda_2}$,  
the Shr\"{o}dinger equation may be written, in a self-explanatory
notation, as
\begin{eqnarray}
  &\big[&\!\!-\frac{\hbar^2}{2 \mu_{\alpha\Lambda_1} }
      \nabla^{2}_{\alpha\Lambda_1}
     -\frac{\hbar^2}{2 \mu_{\alpha\Lambda_2} }
      \nabla^{2}_{\alpha\Lambda_2}
     -\frac{\hbar^2}{ m_\alpha }
      \nabla_{\alpha\Lambda_1}\cdot 
     \nabla_{\alpha\Lambda_1}  \nonumber  \\
      &+&\!\! V_{\alpha \Lambda_1}\!+\! 
       V_{\alpha \Lambda_2}+ V_{\Lambda_1 \Lambda_2}
     - E \; \big]\Psi_{JM}(^6_{\Lambda \Lambda}{\rm He}) = 0 .
\end{eqnarray}
If the third term of the kinetic energy, 
the so-called mass-polarization term,
and $V_{\Lambda_1 \Lambda_2}$ are neglected, we have the trivial 
solution $-E (=B_{\Lambda \Lambda}) = 2 B_\Lambda$. 
Therefore, the quantity 
$\Delta B_{\Lambda \Lambda}=
B_{\Lambda \Lambda}-2B_{\Lambda}$ stands for the contribution
from the  neglected two terms.
In $_\Lambda^{ }$$_\Lambda^6$He, 
the contribution to $B_{\Lambda \Lambda}$ from 
the mass-polarization term
is $+0.13$ MeV which explains the difference between 
$\Delta B_{\Lambda \Lambda}= 1.01$ MeV and
the $\Lambda \Lambda$ bond energy 
${\cal V}^{\rm bond}_{\Lambda \Lambda}=0.88 $ MeV
in Table V. 
This contribution decreases rapidly as the core-nuclear  mass increases
($+0.01$ MeV in $_\Lambda^{ }$$_\Lambda^{\!\!\!10}$Be).

Next, we discuss about  the second
problem, an effect of the $\Lambda N$ spin-spin interaction 
on $\Delta B_{\Lambda \Lambda}$ of Eq.(4.2).
In Fig. 10, the calculated values of 
$\Delta B_{\Lambda \Lambda}$  
are illustrated by the dashed bars.
One notices clearly that  
$\Delta B_{\Lambda \Lambda}$ has 
peculiar mass dependence 
in which some interesting mechanism is included.
It should be remarked here, however, that,
as was already pointed out by Danysz {\it et al.} \cite{Danysz63}, 
the traditional definition of
Eq.(\ref{eq:DeltaBLL}) is of simple meaning only when the
nuclear core is spinless.
On the other hand,
in the case of nuclear core with spin,
the single-$\Lambda$ binding energy $B_\Lambda$ to be
subtracted from 
$B_{\Lambda \Lambda}$ is 
distributed over the ground-state doublet
of the corresponding single-$\Lambda$ hypernucleus.

Here, we remark the fact that the $\Lambda N$ 
spin-spin interaction is not effective (cancelled out) in the
double-$\Lambda$ hypernuclei having the
$\Lambda \Lambda$ spin-singlet pairs.
In the parent single-$\Lambda$ hypernuclei, 
however, the spin-spin interaction
plays an important role
in giving rise to the energy splitting of the
ground-state doublet.
The typical and unique example known experimentally is the
spin-doublet in
$^7_{\Lambda}{\rm Li}$
with $J=\frac{1}{2}^+$(ground; $B_{\Lambda}=5.58$ MeV)
and $J=\frac{3}{2}^+ (E_{\rm x}=0.69$ MeV; $B_{\Lambda}=4.49$ MeV).
Considering this effect, one should use the
spin-averaged value
$\bar{B}_{\Lambda}(^7_{\Lambda}$Li)
$=\frac{1}{3}B_{\Lambda}(\frac{1}{2}^+_{\rm g.s.})
+\frac{2}{3}B_{\Lambda}(\frac{3}{2}^+)$
instead of $B_{\Lambda}(\frac{1}{2}^+_{\rm g.s.}$)
when one likes to deduce $\Delta B_{\Lambda \Lambda}$
from the $_\Lambda^{ }$$_\Lambda^8$Li$(1^+)$ ground state data, if any.
If we adopt this prescription also for the adjacent systems,
we may use 
\vskip 0.3 cm
\noindent
\hskip 0.2 cm
$\bar{B}_{\Lambda}(^6_{\Lambda}{\rm He})=
\frac{1}{4}B_{\Lambda}(1^-_{\rm g.s.})+
\frac{3}{4}B_{\Lambda}(2^-)=4.02$ MeV, 
\vskip 0.2 cm

\noindent
\hskip 0.2 cm
$\bar{B}_{\Lambda}(^6_{\Lambda}{\rm Li})=
\frac{1}{4}B_{\Lambda}(1^-_{\rm g.s.})+
\frac{3}{4}B_{\Lambda}(2^-)=4.31$ MeV, 
\vskip 0.2 cm

\noindent
\hskip 0.2 cm
$\bar{B}_{\Lambda}(^7_{\Lambda}{\rm Li})=
\frac{1}{3}B_{\Lambda}(\frac{1}{2}^+_{\rm g.s.})+
\frac{2}{3}B_{\Lambda}(\frac{3}{2}^+)=5.12$ MeV,  
\vskip 0.2 cm

\noindent
\hskip 0.2 cm
$\bar{B}_{\Lambda}(^8_{\Lambda}{\rm Li})=
\frac{1}{4}B_{\Lambda}(1^-_{\rm g.s.})+
\frac{3}{4}B_{\Lambda}(2^-)=6.58$ MeV,  
\vskip 0.2 cm

\noindent
\hskip 0.2 cm
$\bar{B}_{\Lambda}(^8_{\Lambda}{\rm Be})=
\frac{1}{4}B_{\Lambda}(1^-_{\rm g.s.})+
\frac{3}{4}B_{\Lambda}(2^-)=6.48$ MeV.
\vskip 0.3 cm

\noindent
Here, $B_\Lambda$ of the excited states are taken from
our calculation.
In general, we have
\begin{eqnarray}
\bar{B}_{\Lambda}(^{A-1}_{\Lambda}{\rm Z}) &=& 
\frac{J_0}{2J_0+1}B_{\Lambda}(^{A-1}_{\Lambda}{\rm Z}; 
J_1=J_0-\frac{1}{2})  \nonumber \\
&+& \frac{J_0+1}{2J_0+1}B_{\Lambda}(^{A-1}_{\Lambda}{\rm Z}; 
J_1=J_0+\frac{1}{2}), \nonumber
\label{eq:DeltaBLL3}
\end{eqnarray}
where $J_1=J_0 \pm \frac{1}{2}$ denote the doublet
spins of the $\alpha+x+ \Lambda$ system, $J_0$ being the
ground-state spin of the $\alpha+x$ nuclear core.
For the two spinless cases ($x=0$ and $\alpha$), needless to say,
$\bar{B}_{\Lambda}(^5_{\Lambda}{\rm He})=
B_{\Lambda}(^5_{\Lambda}{\rm He};\frac{1}{2}^+_{\rm g.s.})$
and 
$\bar{B}_{\Lambda}(^9_{\Lambda}{\rm Be})=B_{\Lambda}
(^9_{\Lambda}{\rm Be};\frac{1}{2}^+_{\rm g.s.})$.

Thus, replacing $B_{\Lambda}$
with $\bar{B}_{\Lambda}$ in Eq.(4.2),
we modify 
$\Delta B_{\Lambda \Lambda}$
by $\Delta \bar{B}_{\Lambda \Lambda}$ as
\begin{equation}
\Delta \bar{B}_{\Lambda \Lambda}(^A_{\Lambda \Lambda}{\rm Z}) \equiv
B_{\Lambda \Lambda}(^A_{\Lambda \Lambda}{\rm Z})
-2\bar{B}_{\Lambda}(^{A-1}_{\Lambda}{\rm Z}).
\label{eq:DeltaBLL1}
\end{equation}
\noindent
In Fig. 11, the  solid bars illustrate
$\Delta \bar{B}_{\Lambda \Lambda}$.
Though $\Delta \bar{B}_{\Lambda \Lambda}$
is free from the effect of
the $\Lambda N$ spin-spin interaction,
its magnitude for $A=7 - 10$ deviates significantly  from 
$\Delta \bar{B}_{\Lambda \Lambda}$
($_\Lambda^{ }$$_\Lambda^6$He)$=1.01$ MeV.
The deviation comes from the effect of the dynamical change 
in the core nucleus structure (shrinkage in the $\alpha-x$ distance) 
due to the partition of the $\Lambda$ hyperons, and turns out 
to be maximum in the case of $_\Lambda^{ }$$_\Lambda^{\!\!\!10}$Be.
We emphasize that, 
even if one  employs $\Delta \bar{B}_{\Lambda \Lambda}$, 
it is impossible to extract
any consistent value of the $\Lambda \Lambda$ bond energy
from Fig. 11 in which $\Delta \bar{B}_{\Lambda \Lambda}$
scatters in a range of a factor of two.

As mentioned above, a consistent estimation of the
$\Lambda \Lambda$ bond energy ($ 0.9 - 1.0 $ MeV, 
nearly independent of the mass number, as seen in Table V)
can be obtained by taking 
${\cal V}^{\rm bond}_{\Lambda \Lambda}$
of Eq. (4.1) 
as the definition of that energy, though help of
the theoretical calculation with $V_{\Lambda \Lambda}=0$  
is necessary.


\section{Summary}

We have carried out structure calculations of
$_\Lambda^{ }$$_\Lambda^6$He, $_\Lambda^{ }$$_\Lambda^7$He,
$_\Lambda^{ }$$_\Lambda^7$Li, $_\Lambda^{ }$$_\Lambda^8$Li,
$_\Lambda^{ }$$_\Lambda^9$Li
, $_\Lambda^{ }$$_\Lambda^9$Be and
$_\Lambda^{ }$$_\Lambda^{\!\!\!10}$Be taking the framework of
$\alpha +x+ \Lambda + \Lambda$  model
with $x=0, n, p, d, t, ^3$He and $\alpha$, respectively.
We determined 
the interactions between  constituent particles
so as to reproduce reasonably
the  observed low-energy properties of the $\alpha +x$ nuclei 
and  the existing data of $\Lambda$-binding energies of the
$x+\Lambda$  and $\alpha+x+\Lambda$ systems.
The $\Lambda \Lambda$ interaction was constructed so as to
reproduce the $B_{\Lambda \Lambda}$($_\Lambda^{ }$$_\Lambda^6$He)
given by the NAGARA event within our $\alpha+\Lambda+\Lambda$ model,
where the long-range part of our interaction was adjusted to
simulate the behavior of the appropriate OBE model (NF).
With no  adjustable parameters, the four-body calculations 
of the $\alpha+x+\Lambda+\Lambda$ systems
were performed accurately 
using the Jacobian-coordinate Gaussian-basis 
coupled-rearrangement-channel method.
Obtained energy spectra of the double-$\Lambda$ hypernuclei
with $A=6 - 10$ are summarized in Fig. 12.

Major results to be emphasized here are as follows:

(1)  It is striking that 
the calculated $B_{\Lambda \Lambda}$ of the $2^+$ excited state in
$_\Lambda^{ }$$_\Lambda^{\!\!\!10}$Be, 12.28 MeV, 
agrees with the experimental value $B_{\Lambda \Lambda}^{\rm exp}$ 
($_\Lambda^{ }$$_\Lambda^{\!\!\!10}$Be)
$=12.33\pm ^{0.35}_{0.21}$ MeV 
in the {\it Demachi-Yanagi} event \cite{Ichi99,Ichi01}.
We therefore interpret this event as observation of the
$2^+$ excited state of $_\Lambda^{ }$$_\Lambda^{\!\!\!10}$Be.
The agreement suggests that our systematic calculations
are predictive for coming
events expected to be found in the
further analysis of the E373 data, {\it etc}.

(2) Together with the energy spectrum of each double-$\Lambda$ 
hypernucleus, those of
the corresponding core nucleus and 
single-$\Lambda$ hypernucleus are exhibited side by side 
in Figs.$2 - 7$ so as to see clearly that
injection of one  and two $\Lambda$ particles 
leads to  stronger binding of the whole system
and prediction of more bound states.
In the  bound states of
any double-$\Lambda$ hypernucleus, 
two $\Lambda$ particles are dominantly coupled  to 
$S=0$ and hence the spin and  parity become the same as 
those of its nuclear core, but 
the theoretical $B_{\Lambda \Lambda}$ values 
are of importance to guide the analysis of the
emulsion experiments.  

(3) Dynamical change of the $\alpha+x$ nuclear core 
by the participation of 
the $\Lambda$ particles is substantially seen in 
double-$\Lambda$ hypernuclei; there occurs, averagely speaking,
about 8 \% shrinkage of the
$\alpha-x$ distance compared with the distance in 
the single-$\Lambda$ hypernucleus.
This shrinkage is realized by the 
large energy gain in the $\Lambda-\alpha$ and $\Lambda-x$ 
parts which overcomes 
the energy loss in the $\alpha-x$ relative motion.

(4) We estimated the $\Lambda \Lambda$ bond energy
using the faithful definition
${\cal V}^{\rm bond}_{\Lambda \Lambda}$
$= B_{\Lambda \Lambda} - 
B_{\Lambda \Lambda}(V_{\Lambda \Lambda}=0)$ and found
it to be 0.88 MeV for $_\Lambda^{ }$$_\Lambda^6$He
and $0.93 -0.98$ MeV for the other 
double-$\Lambda$ hypernuclei.
We demonstrated that
the quantity  
$\Delta B_{\Lambda \Lambda}=B_{\Lambda \Lambda} -2B_\Lambda$  
is not a good measure of the $\Lambda \Lambda$
bond energy since 
$\Delta B_{\Lambda \Lambda}$ is free from neither 
the contribution from the splitting of the
ground-state doublet in the single-$\Lambda$ hypernucleus
nor that of the structure change of the core nucleus.
In fact, the value of $\Delta B_{\Lambda \Lambda}$ 
scatters from 0.28 to 1.68 MeV for the
double-$\Lambda$ hypernuclei with $A=6 - 10$. 
We then modified $\Delta B_{\Lambda \Lambda}$ 
by $\Delta \bar{B}_{\Lambda \Lambda}=B_{\Lambda \Lambda}
-2\bar{B}_\Lambda$ with $\bar{B}_\Lambda$ being the spin-average of
$B_\Lambda$'s for the ground-state spin-doublet. We found,
however,  that
$\Delta \bar{B}_{\Lambda \Lambda}$ still ranges from 0.83
to 1.68 MeV due to the  structure change of the core nucleus.
Direct use of $B_{\Lambda \Lambda}$ itself rather than the 
use of  $\Delta B_{\Lambda \Lambda}$ or
$\Delta \bar{B}_{\Lambda \Lambda}$ is recommended
when the experimental result  and calculational result
are compared to each other.

In conclusion, the present precise and extensive four-body cluster-model
calculation can be an opening of the spectroscopic study of
double-$\Lambda$ hypernuclei.

%
%
\section*{Acknowledgments}

The authors would like to thank  Professor K. Nakazawa
and Dr. H. Takahashi for valuable discussions and information on 
the experimental project KEK-E373.
They are also thankful to Professor Y. Akaishi, 
Professor K. Ikeda and Professor A. Gal
for helpful discussions and encouragement.
One of the authors (T.M.) thanks S. Kahana, L. McLerran,
D.J. Millener and Physics Department of Brookhaven National
Laboratory for their hospitality and support. He is also
grateful to the Institute for Nuclear Theory at the
University of Washington for its hospitality.
This work was supported  
by the Grant-in-Aid for Scientific Research 
of Monbukagakushou of Japan.


\vfill\eject


\noindent
TABLE I.  Parameters of (a)  $\alpha \alpha$ interaction,
(b)  $\alpha \, t \:(\Lambda ^3$He) interaction,
(c)  $\alpha \,d$ interaction
and (d)  $\alpha \, N$ interaction
defined in Eq.(3.2). Size parameters  are 
in fm$^{-2}$ and strengths are in MeV.
The $^1S_0$ scattering length is -0.575 fm 
    and the effective range is 6.45 fm.
$$\vbox{
\offinterlineskip
\halign{
%
%
        \enspace\hfil#\hfil\enspace &
        \enspace\hfil#\hfil\enspace &
        \enspace\hfil#\hfil\enspace &
        \enspace\hfil#\hfil\enspace \cr
%
%
\noalign{\hrule height 0.6pt}
\noalign{\vskip 0.07 true cm} \cr
\noalign{\hrule height 0.6pt}
\noalign{\vskip 0.20 true cm} \cr
&\multispan2 (a) $\alpha \,\alpha$ interaction &  \cr
\noalign{\vskip 0.20 true cm} \cr
 $i$  &1 & 2 & 3 \cr
\noalign{\vskip 0.20 true cm} \cr
 $\beta_i$   &0.1111  &0.2777  &  0.3309    \cr
\noalign{\vskip 0.15 true cm} \cr
  $V_i$  &$-1.742$  &$-395.9$  & 299.4   \cr
\noalign{\vskip 0.15 true cm} \cr
  $V_i^{\rm p}$  &$ 0.0 $  &$ 0.0 $  & 0.0   \cr
\noalign{\vskip 0.15 true cm} \cr
\noalign{\hrule height 0.6pt}
\noalign{\vskip 0.20 true cm} \cr
&\multispan2 (b)  $\alpha \, t \:(\alpha ^3$He) interaction
 &  \cr
\noalign{\vskip 0.20 true cm} \cr
 $i$  &1  &2  &3  \cr
\noalign{\vskip 0.20 true cm} \cr
 $\beta_i$   &0.0913 &0.1644  &0.2009    \cr
\noalign{\vskip 0.10 true cm} \cr
  $V_i$  &6.9  &$-43.35$  &$-51.7$  \cr
\noalign{\vskip 0.10 true cm} \cr
 $\beta_i^{\rm p}$   &0.0913 &0.1644  &0.2009    \cr
\noalign{\vskip 0.10 true cm} \cr
  $V_i^{\rm p}$  &6.9 &43.35  &$-51.7$  \cr
\noalign{\vskip 0.20 true cm} \cr
 $\gamma_i$     &  0.28    \cr
\noalign{\vskip 0.10 true cm} \cr
  $V_i^{ls}$  & $-1.2$   \cr
\noalign{\vskip 0.10 true cm} \cr
 $\gamma_i^{\rm p}$     &  0.28    \cr
\noalign{\vskip 0.10 true cm} \cr
  $V_i^{ls,{\rm p}}$  & 1.2   \cr
\noalign{\vskip 0.15 true cm} \cr
\noalign{\hrule height 0.6pt}
\noalign{\vskip 0.20 true cm} \cr
&\multispan2 (c) $\alpha \,d$ interaction &  \cr
\noalign{\vskip 0.20 true cm} \cr
 $i$  &1  \cr
\noalign{\vskip 0.20 true cm} \cr
 $\beta_i$     &  0.2   \cr
\noalign{\vskip 0.10 true cm} \cr
  $V_i$  & $-64.21$  \cr
\noalign{\vskip 0.10 true cm} \cr
 $\beta_i^{\rm p}$     &  0.2   \cr
\noalign{\vskip 0.10 true cm} \cr
  $V_i^{\rm p}$  & $-10.21$  \cr
\noalign{\vskip 0.20 true cm} \cr
 $\gamma_i$     & 0.3   \cr
\noalign{\vskip 0.10 true cm} \cr
  $V_i^{ls}$  & $-4.0$   \cr
\noalign{\vskip 0.10 true cm} \cr
 $\gamma_i^{\rm p}$     &  0.3   \cr
\noalign{\vskip 0.15 true cm} \cr
  $V_i^{ls,{\rm p}}$  & $-4.0$   \cr
\noalign{\vskip 0.15 true cm} \cr
\noalign{\hrule height 0.6pt}
\noalign{\vskip 0.20 true cm} \cr
&\multispan2 (d) $\alpha \,N$ interaction &  \cr
\noalign{\vskip 0.20 true cm} \cr
 $i$  &1 & 2 & 3 \cr
\noalign{\vskip 0.20 true cm} \cr
 $\beta_i$    &0.36  &  0.9    \cr
\noalign{\vskip 0.10 true cm} \cr
  $V_i$  &$-96.3$   &77.0   \cr
\noalign{\vskip 0.10 true cm} \cr
 $\beta_i^{\rm p}$     &  0.2  & 0.53 &2.5   \cr
\noalign{\vskip 0.10 true cm} \cr
  $V_i^{\rm p}$  & 34.0 &$-85.0$  &51.0  \cr
\noalign{\vskip 0.20 true cm} \cr
 $\gamma_i$   &0.396   &0.52  &2.2   \cr
\noalign{\vskip 0.10 true cm} \cr
  $V_i^{ls}$  &$-20.0$ &$-16.8$  &20.0   \cr
\noalign{\vskip 0.10 true cm} \cr
 $\gamma_i^{\rm p}$   &0.396    &2.2   \cr
\noalign{\vskip 0.10 true cm} \cr
  $V_i^{ls,{\rm p}}$  & 6.0 & $-6.0$   \cr
\noalign{\vskip 0.15 true cm} \cr
\noalign{\hrule height 0.6pt}
\noalign{\vskip 0.07 true cm} \cr
\noalign{\hrule height 0.6pt}
}}$$


\vskip 1.0 true cm

\noindent
Table II.  Parameters of (a)  $\Lambda \,\alpha$ interaction,
(b)  $\Lambda \, t \:(\Lambda ^3$He) interaction and
(c)  $\Lambda \,d$ interaction 
defined in Eq.(3.4). Size parameters  are 
in fm$^{-2}$ and strengths are in MeV.
$$\vbox{
\offinterlineskip
\halign{
%
%
        \enspace\hfil#\hfil\enspace &
        \enspace\hfil#\hfil\enspace &
        \enspace\hfil#\hfil\enspace &
        \enspace\hfil#\hfil\enspace \cr
%
%
\noalign{\hrule height 0.6pt}
\noalign{\vskip 0.07 true cm} \cr
\noalign{\hrule height 0.6pt}
\noalign{\vskip 0.20 true cm} \cr
&\multispan2 (a) $\Lambda \,\alpha$ interaction &  \cr
\noalign{\vskip 0.20 true cm} \cr
 $i$  &1 & 2 & 3 \cr
\noalign{\vskip 0.2 true cm} \cr
 $\beta_i$     &  0.2752  & 0.4559  & 0.6123   \cr
\noalign{\vskip 0.10 true cm} \cr
  $V_i$  & $-17.49$ & $-127.0$  & 497.8  \cr
\noalign{\vskip 0.10 true cm} \cr
  $V_i^{\rm s} $  &0.0 &0.0  &0.0  \cr
\noalign{\vskip 0.20 true cm} \cr
 $\gamma_i$     & 0.1808  & 0.1808 & 0.1808   \cr
\noalign{\vskip 0.10 true cm} \cr
 $\delta_i$    &  0.4013 & 0.9633  & 2.930   \cr
\noalign{\vskip 0.10 true cm} \cr
  $U_i$  & $-0.3706$ & $-12.94$ & $-331.2$  \cr
\noalign{\vskip 0.10 true cm} \cr
  $U_i^{\rm s} $  & 0.0 & 0.0  & 0.0  \cr
\noalign{\vskip 0.15 true cm} \cr
\noalign{\hrule height 0.6pt}
\noalign{\vskip 0.20 true cm} \cr
&\multispan2 (b) $\Lambda \, t \:(\Lambda  ^3$He) 
interaction &\cr
\noalign{\vskip 0.20 true cm} \cr
 $i$  &1 & 2 & 3 \cr
\noalign{\vskip 0.20 true cm} \cr
 $\beta_i$    &  0.2874  & 0.4903  &0.6759   \cr
\noalign{\vskip 0.15 true cm} \cr
  $V_i$  & $-14.16$ & $-108.0$  & 425.9  \cr
\noalign{\vskip 0.10 true cm} \cr
  $V_i^{\rm s} $  & 2.379 & 10.91  & $-126.9$  \cr
\noalign{\vskip 0.20 true cm} \cr
 $\gamma_i$     &  0.2033  & 0.2033  & 0.2033   \cr
\noalign{\vskip 0.10 true cm} \cr
 $\delta_i$    &  0.3383  & 0.8234  & 2.521   \cr
\noalign{\vskip 0.10 true cm} \cr
  $U_i$  & $-0.2701 $ & $-9.553$  & $-231.6$  \cr
\noalign{\vskip 0.10 true cm} \cr
  $U_i^{\rm s} $  & $-0.2615$ & 1.433  & 97.05  \cr
\noalign{\vskip 0.15 true cm} \cr
\noalign{\hrule height 0.6pt}
\noalign{\vskip 0.20 true cm} \cr
&\multispan2 (c) $\Lambda \,d$ interaction &  \cr
\noalign{\vskip 0.20 true cm} \cr
 $i$  &1 & 2 & 3 \cr
\noalign{\vskip 0.20 true cm} \cr
 $\beta_i$     &0.3153 & 0.5773  & 0.8532   \cr
\noalign{\vskip 0.10 true cm} \cr
  $V_i$  & $-10.84$ & $-88.36$  & 167.2  \cr
\noalign{\vskip 0.10 true cm} \cr
  $V_i^{\rm s} $  & 2.734 & 14.35  & $-179.9$  \cr
\noalign{\vskip 0.20 true cm} \cr
 $\gamma_i$     &  0.2710  & 0.2710  & 0.2710   \cr
\noalign{\vskip 0.10 true cm} \cr
 $\delta_i$    &  0.2470  & 0.4870  & 1.924   \cr
\noalign{\vskip 0.10 true cm} \cr
  $U_i$  & $-0.1862$ & $-5.844$  & $-3.065$  \cr
\noalign{\vskip 0.10 true cm} \cr
  $U_i^{\rm s} $  & $-0.2705$ & 1.566  & 100.4  \cr
\noalign{\vskip 0.15 true cm} \cr
\noalign{\hrule height 0.6pt}
\noalign{\vskip 0.07 true cm} \cr
\noalign{\hrule height 0.6pt}
}}$$


\vskip 1.0 true cm

\noindent
TABLE III.  Parameters of the $\Lambda \,N$ interaction
defined in Eq.(3.5) which is used only in the 
$\alpha+N+\Lambda$ and $\alpha+N+\Lambda+\Lambda$ systems $(x=N)$.
Size parameters  are 
in fm$^{-2}$  and strengths are in MeV.
$$\vbox{
\offinterlineskip
\halign{
%
%
        \enspace\hfil#\hfil\enspace &
        \enspace\hfil#\hfil\enspace &
        \enspace\hfil#\hfil\enspace &
        \enspace\hfil#\hfil\enspace \cr
%
%
\noalign{\hrule height 0.6pt}
\noalign{\vskip 0.07 true cm} \cr
\noalign{\hrule height 0.6pt}
\noalign{\vskip 0.20 true cm} \cr
&\multispan2  $\Lambda \,N$ interaction when $x=N$ &  \cr
\noalign{\vskip 0.20 true cm} \cr
 $i$  &1 & 2 & 3 \cr
\noalign{\vskip 0.20 true cm} \cr
 $\mu_i$     &0.5487  & 1.384  & 6.250   \cr
\noalign{\vskip 0.10 true cm} \cr
  $v_i^{\rm even}$  & $-10.40$ & $-87.05$  &1031  \cr
\noalign{\vskip 0.10 true cm} \cr
  $v_i^{{\rm even}, \sigma}$  &0.2574 &17.09  &$-256.3$  \cr
\noalign{\vskip 0.10 true cm} \cr
  $v_i^{\rm odd}$  &$-5.816$ & $-18.29$  & 4029  \cr
\noalign{\vskip 0.10 true cm} \cr
  $v_i^{{\rm odd}, \sigma}$ & $-0.959$ & $-9.184$  & $-573.8$  \cr
\noalign{\vskip 0.15 true cm} \cr
\noalign{\hrule height 0.6pt}
\noalign{\vskip 0.07 true cm} \cr
\noalign{\hrule height 0.6pt}
}}$$


\vskip 1.0 true cm

\noindent
TABLE IV.  Parameters of the $\Lambda \,\Lambda$ interaction
defined in Eq.(3.6).
Size parameters  are 
in fm$^{-2}$  and strengths are in MeV.
 The $^1S_0$ scattering length is -0.575 fm 
    and the effective range is 6.45 fm.
$$\vbox{
\offinterlineskip
\halign{
%
%
        \enspace\hfil#\hfil\enspace &
        \enspace\hfil#\hfil\enspace &
        \enspace\hfil#\hfil\enspace &
        \enspace\hfil#\hfil\enspace \cr
%
%
\noalign{\hrule height 0.6pt}
\noalign{\vskip 0.07 true cm} \cr
\noalign{\hrule height 0.6pt}
\noalign{\vskip 0.20 true cm} \cr
&\multispan2  $\Lambda \,\Lambda$ interaction &  \cr
\noalign{\vskip 0.20 true cm} \cr
 $i$  &1 & 2 & 3 \cr
\noalign{\vskip 0.20 true cm} \cr
 $\mu_i$     & 0.555  & 1.656  & 8.163   \cr
\noalign{\vskip 0.10 true cm} \cr
  $v_i$  &$-10.67$ & $-93.51$  & 4884  \cr
\noalign{\vskip 0.10 true cm} \cr
  $v_i^{\sigma}$  &0.0966  & 16.08  & 915.8  \cr
\noalign{\vskip 0.15 true cm} \cr
\noalign{\hrule height 0.6pt}
\noalign{\vskip 0.07 true cm} \cr
\noalign{\hrule height 0.6pt}
}}$$


\vskip 1.0 true cm

\noindent
Table V.  Calculated energies of the ground states of
          $A=6 - 10$ double-$\Lambda$ hypernuclei
          based on the $\alpha + x +\Lambda+ \Lambda$ 
          four-body model 
          $(x=0, n, p, d, t, ^3{\rm He},$ and $\alpha)$.
          $E_{\Lambda \Lambda}$ are measured
          from the $\alpha + x +\Lambda+ \Lambda$ threshold. 
          The $\Lambda \Lambda$ bond energy 
          ${\cal V}^{\rm bond}_{\Lambda \Lambda}$  is defined by 
          Eq.(4.1).
          Information on the $2^+$ excited state of
          $_\Lambda^{ }$$_\Lambda^{\!\!\!10}$Be
          is specially added so as to demonstrate the agreement with 
          the experimental result.

$$\vbox{
\offinterlineskip
\halign{
%
%
        \enspace\hfil#\hfil\enspace &
        \enspace\hfil#\hfil\enspace &
        \enspace\hfil#\hfil\enspace &
        \enspace\hfil#\hfil\enspace &
        \enspace\hfil#\hfil\enspace &
        \enspace\hfil#\hfil\enspace \cr
%
%
\noalign{\hrule height 0.6pt}
\noalign{\vskip 0.05 true cm} \cr
\noalign{\hrule height 0.6pt}
\noalign{\vskip 0.25 true cm} \cr
  &$J^{\pi}$  &$E_{\Lambda \Lambda}$  &$B_{\Lambda \Lambda}$  
&$B_{\Lambda \Lambda}^{\rm exp}$ 
&${\cal V}^{\rm bond}_{\Lambda \Lambda}$ \cr
 &  &(MeV)  &(MeV)  &(MeV)  &(MeV)    \cr
\noalign{\vskip 0.25 true cm} \cr
\noalign{\hrule}
\noalign{\vskip 0.25 true cm} \cr
$_\Lambda^{ }$$_\Lambda^6$He  &$0^+$  &$-7.25$  &7.25  
&$7.25 \pm 0.19\,^a$   &0.88   \cr
\noalign{\vskip 0.25 true cm} \cr
$_\Lambda^{ }$$_\Lambda^7$He  &$\frac{3}{2}^-$  &$-8.47$  &9.36  
&- &0.96   \cr
\noalign{\vskip 0.25 true cm} \cr
$_\Lambda^{ }$$_\Lambda^7$Li  &$\frac{3}{2}^-$  &$-7.48$  &9.45 &-  
&0.95 \cr
\noalign{\vskip 0.25 true cm} \cr
$_\Lambda^{ }$$_\Lambda^8$Li  &$1^+$  &$-12.10$   &11.44  &- &0.98 \cr
\noalign{\vskip 0.25 true cm} \cr
$_\Lambda^{ }$$_\Lambda^9$Li  &$\frac{3}{2}^-$  &$-17.05$  &14.55  &-  
&0.98 \cr
\noalign{\vskip 0.25 true cm} \cr
$_\Lambda^{ }$$_\Lambda^9$Be  &$\frac{3}{2}^-$  &$-16.00$  &14.40  &-  
 & 0.97   \cr
\noalign{\vskip 0.25 true cm} \cr
$_\Lambda^{ }$$_\Lambda^{\!\!\!10}$Be  &$0^+$  &$-15.05$  &15.14  
&$17.7\pm0.4\,^b $ &0.93   \cr
\noalign{\vskip 0.05 true cm} \cr
  &  &  &  
&$14.6\pm0.4\,^b $ &    \cr
\noalign{\vskip 0.25 true cm} \cr
$_\Lambda^{ }$$_\Lambda^{\!\!\!10}$Be  &$2^+$  &$-12.19$  &12.28  
&$12.33{\pm^{0.35}_{0.21}}\,^c$  &0.93   \cr
\noalign{\vskip 0.25 true cm} \cr
\noalign{\hrule height 0.6pt}
\noalign{\vskip 0.05 true cm} \cr
\noalign{\hrule height 0.6pt}
}}$$

$^a$ Ref.\cite{Nagara}.

$^b$ Ref.\cite{Danysz63}. Also see text for the second value.

$^c$ Ref.\cite{Ichi99,Ichi01}.


\vskip 1.0 true cm

\noindent
Table VI. Calculated r.m.s. 
          distances between  $\alpha$ and $x$, 
          $\bar{r}_{\alpha x}$, in core nuclei, single 
          $\Lambda$ hypernuclei and double-$\Lambda$ hypernuclei
          ($x=n, d, t, \alpha$).
          The expectation values of kinetic energy and potential energy
          between $\alpha$ and $x$, $<T_{\alpha x}>$,
          $<V_{\alpha  x}>$ and 
          $<T_{\alpha  x}+V_{\alpha  x}>$  are also listed.
          For  $^5$He and $^8$Be, $\bar{r}_{\alpha -x}$ are not 
         calculated since they are resonant states.

$$\vbox{
\offinterlineskip
\halign{
%
%
        \enspace\hfil#\hfil\enspace &
        \enspace\hfil#\hfil\enspace &
        \enspace\hfil#\hfil\enspace &
        \enspace\hfil#\hfil\enspace &
        \enspace\hfil#\hfil\enspace &
        \enspace\hfil#\hfil\enspace &
        \enspace\hfil#\hfil\enspace &
        \enspace\hfil#\hfil\enspace \cr
%
%
\noalign{\hrule height 0.6pt}
\noalign{\vskip 0.05 true cm} \cr
\noalign{\hrule height 0.6pt}
\noalign{\vskip 0.25 true cm} \cr
     &$\bar{r}_{\alpha  x}$ &$<\!T_{\alpha x}\!>$
    &$<\!V_{\alpha  x}\!>$  &$<\!T_{\alpha x}\!+\!V_{\alpha  x}\!>$
\cr
\noalign{\vskip 0.25 true cm} \cr
\noalign{\hrule}
\noalign{\vskip 0.25 true cm} \cr
$^5$He  &$-$  &7.86  &$-6.97$  &0.89 \cr
\noalign{\vskip 0.25 true cm} \cr
$^6_{\Lambda}$He  &5.79  &11.38  &$-9.92$  &1.46 \cr
\noalign{\vskip 0.25 true cm} \cr
$_\Lambda^{ }$$_\Lambda^7$He  &3.92   &15.19  &$-11.95$  &2.24 \cr
\noalign{\vskip 0.25 true cm} \cr
\noalign{\hrule}
\noalign{\vskip 0.25 true cm} \cr
$^6$Li  &4.10   &11.59  &$-13.06$  &$-1.47$ \cr
\noalign{\vskip 0.25 true cm} \cr
$^7_{\Lambda}$Li  &3.44  &15.59  &$-16.70$  &$-1.11$ \cr
\noalign{\vskip 0.25 true cm} \cr
$_\Lambda^{ }$$_\Lambda^8$Li  &3.16  &18.86  &$-19.54$  &$-0.68$ \cr
\noalign{\vskip 0.25 true cm} \cr
\noalign{\hrule}
\noalign{\vskip 0.25 true cm} \cr
$^7$Li  &3.69   &17.45  &$-19.95$  &$-2.50$ \cr
\noalign{\vskip 0.25 true cm} \cr
$^8_{\Lambda}$Li  &3.30  &21.85  &$-24.00$  &$-2.15$ \cr
\noalign{\vskip 0.25 true cm} \cr
$_\Lambda^{ }$$_\Lambda^9$Li  &3.05  &26.74  &$-28.33$  &$-1.59$ \cr
\noalign{\vskip 0.25 true cm} \cr
\noalign{\hrule}
\noalign{\vskip 0.25 true cm} \cr
$^8$Be  &$-$    &7.21  &$-7.12$  &0.09 \cr
\noalign{\vskip 0.25 true cm} \cr
$^9_{\Lambda}$Be  &3.78  &14.90  &$-14.14$  &0.76 \cr
\noalign{\vskip 0.25 true cm} \cr
$_\Lambda^{ }$$_\Lambda^{\!\!\!10}$Be  &3.44  &19.49  
&$-17.96$  &1.53 \cr
\noalign{\vskip 0.25 true cm} \cr
\noalign{\hrule height 0.6pt}
\noalign{\vskip 0.05 true cm} \cr
\noalign{\hrule height 0.6pt}
}}$$


\vskip 1.8 cm

{\bf Figure captions} 

\vskip 0.8 cm
FIG. 1. Jacobian coordinates for all the rearrangement 
channels ($c=1-9$) of the $\alpha+x+\Lambda+\Lambda$ four-body system. 
Two $\Lambda$ particles
are to be antisymmetrized, and  $\alpha$ and $x$ are to 
be symmetrized when $x=\alpha$.

\vskip 0.5 cm
FIG. 2. Calculated energy levels of $~^5{\rm He},
~^6_{\Lambda}{\rm He}$ and 
$_\Lambda^{ }$$_\Lambda^7$He on the basis of
the $\alpha+n$, 
$\alpha+n+\Lambda$ and $\alpha+n+\Lambda+\Lambda$ 
models, respectively. The level energies are measured from the
particle break-up thresholds.

\vskip 0.5 cm
FIG. 3. Calculated energy levels of $~^5{\rm Li},
~^6_{\Lambda}{\rm Li}$ and 
$_\Lambda^{ }$$_\Lambda^7$Li on the basis of
the $\alpha+p$, 
$\alpha+p+\Lambda$ and $\alpha+p+\Lambda+\Lambda$ 
models, respectively. The level energies are measured from the
particle break-up thresholds.

\vskip 0.5 cm
FIG. 4. Calculated energy levels of $~^6{\rm Li},
~^7_{\Lambda}{\rm Li}$ and 
$_\Lambda^{ }$$_\Lambda^8$Li on the basis of
the $\alpha+d$, 
$\alpha+d+\Lambda$ and $\alpha+d+\Lambda+\Lambda$ 
models, respectively. The level energies are measured from the
particle break-up thresholds or are given 
by excitation energies $E_{\rm x}$.

\vskip 0.5 cm
FIG. 5. Calculated energy levels of $~^7{\rm Li},
~^8_{\Lambda}{\rm Li}$ and 
$_\Lambda^{ }$$_\Lambda^9$Li on the basis of
the $\alpha+t$, 
$\alpha+t+\Lambda$ and $\alpha+t+\Lambda+\Lambda$ 
models, respectively. The level energies are measured from the
particle break-up thresholds or are given 
by  excitation energies $E_{\rm x}$.

\vskip 0.5 cm
FIG. 6. Calculated energy levels of $~^7{\rm Be},
~^8_{\Lambda}{\rm Be}$ and 
$_\Lambda^{ }$$_\Lambda^9$Be on the basis of
the $\alpha+~^3{\rm He}$, 
$\alpha+~^3{\rm He}+\Lambda$ 
and $\alpha+~^3{\rm He}+\Lambda+\Lambda$ 
models, respectively. The level energies are measured from the
particle break-up thresholds or are given 
by excitation energies $E_{\rm x}$.

\vskip 0.5 cm
FIG. 7. Calculated energy levels of $~^8{\rm Be},
~^9_{\Lambda}{\rm Be}$ and 
$_\Lambda^{ }$$_\Lambda^{\!\!\!10}$Be 
on the basis of the $\alpha+\alpha$, 
$\alpha+\alpha+\Lambda$ and $\alpha+\alpha+\Lambda+\Lambda$ 
models, respectively. The level energies are measured from the
particle break-up thresholds or are given 
by excitation energies $E_{\rm x}$.

\vskip 0.5 cm

FIG. 8. The $\alpha-n$ two-body densities (correlation functions),
$\rho(r_{\alpha n}$), of $^5$He($3/2^-$), $^6_{\Lambda}$He($1^-$) and 
$_\Lambda^{ }$$_\Lambda^7$He($3/2^-$).
Here, it is multiplied by $r_{\alpha n}^2$.

\vskip 0.5 cm

FIG. 9. Calculated values of 
$B_{\Lambda \Lambda}(^A_{\Lambda \Lambda}Z)$
in the ground states are given by closed circles.
The same quantities but calculated 
by putting $V_{\Lambda \Lambda}=0$,
namely $B_{\Lambda \Lambda}(^A_{\Lambda \Lambda}$Z;
$V_{\Lambda \Lambda}=0$),
are shown by open circles.

\vskip 0.5 cm
FIG. 10. Calculated values of 
$\Delta B_{\Lambda \Lambda}(^A_{\Lambda \Lambda}Z)$
defined in Eq. (4.2). 

\vskip 0.5 cm
FIG. 11. Calculated values of 
$\Delta {\bar B}_{\Lambda \Lambda}(^A_{\Lambda \Lambda}Z)$
defined in Eq. (4.3). 

\vskip 0.5 cm
FIG. 12. Summary of the energy levels of
the double-$\Lambda$ hypernuclei
$_\Lambda^{ }$$_\Lambda^6$He, 
$_\Lambda^{ }$$_\Lambda^7$He, 
$_\Lambda^{ }$$_\Lambda^7$Li, 
$_\Lambda^{ }$$_\Lambda^8$Li, 
$_\Lambda^{ }$$_\Lambda^9$Li, 
$_\Lambda^{ }$$_\Lambda^9$Be and 
$_\Lambda^{ }$$_\Lambda^{\!\!\!10}$Be
calculated using the $\alpha+x+\Lambda+\Lambda$ model
with $x=0, n, p, d, t, ^3$He and $\alpha$, respectively.


\begin{thebibliography}{99}

\bibitem{Nagara} H. Takahashi {\it et al.}, Phys. Rev. Letters 
{\bf 87}, 212502 (2001).

\bibitem{Danysz63} M. Danysz {\it et al.}, Nucl. Phys. 
{\bf 49}, 121 (1963).

\bibitem{Prowse66} D.J. Prowse, Phys. Rev. Letters 
{\bf 17}, 782 (1966).

\bibitem{Dalitz89} R.H. Dalitz, D.H. Davis, P.H. Fowler, A. Montwill,
  J. Pniewski, and J.A. Zakrzewski, Proc. Roy. Soc. Lond.
  {\bf A426}, 1 (1989).

\bibitem{Aoki91} S. Aoki {\it et al.}, Prog. Theor. Phys. {\bf 85},
1287 (1991).

\bibitem{Dover91}C.B. Dover, D.J. Millener, A. Gal, and D.H. Davis,
Phys. Rev. C44, 1905 (1991).

\bibitem{Yamamoto91}Y. Yamamoto, H. Takaki, and K. Ikeda, 
Prog. Theor. Phys. {\bf 86}, 867 (1991). 

\bibitem{Takaki89} H. Takaki, Wang Xi-cang, and H. Band\={o}, Prog. 
Theor. Phys. {\bf 83}, 13 (1989).

\bibitem{Bodmer84} A.R. Bodmer, Q.N. Usmani, and J. Carlson, 
 Nucl. Phys. {\bf A422}, 510 (1984).

\bibitem{Bodmer87} A.R. Bodmer and Q.N. Usmani, 
 Nucl. Phys. {\bf A468}, 653 (1987).

\bibitem{Gal01}I.N. Filikhin and A. Gal, 
Phys. Rev. {\bf C65}, 041001 (2002).

\bibitem{Gal02}I.N. Filikhin and A. Gal, nucl-th.0203036 (2002).

\bibitem{Hiyama97} E. Hiyama, M. Kamimura, T. Motoba, T. Yamada, and 
Y. Yamamoto, Prog. Theor. Phys. {\bf 97}, 881 (1997). 

\bibitem{Furutani80} H. Furutani, H. Kanada, T. Kaneko,
S. Nagata, H. Nishioka, S. Okabe, S. Saito, T. Sakuda, and M. Seya,
Prog. Thore. Phys. Suppl. No. 68, 193 (1980).

\bibitem{Kami88} M. Kamimura, Phys. Rev. {\bf A38}, 621 (1988).

\bibitem{Kame89} H. Kameyama, M. Kamimura, and Y. Fukushima,
Phys. Rev. {\bf C40}, 974 (1989).

\bibitem{Hiyama95} E. Hiyama and M. Kamimura, Nucl. Phys.
{\bf A588}, 35c (1995).

\bibitem{Hiya96} E. Hiyama, M. Kamimura, T. Motoba,
T. Yamada, and Y. Yamamoto, Phys. Rev. {\bf C53}, 2075 (1996).




\bibitem{Hiyama00} E. Hiyama, M. Kamimura, 
T. Motoba, T. Yamada, and Y. Yamamoto,
Phy. Rev. Letters {\bf 85}, 270 (2000).


\bibitem{Hiyama02} E. Hiyama, M. Kamimura, 
T. Motoba, T. Yamada, and Y. Yamamoto,
Phys. Rev. {\bf C64}, 011301 (2002).


\bibitem{Kamada01} H. Kamada, A. Nogga, W. Gloeckle, 
E. Hiyama, M. Kamimura, K. Varga, Y. Suzuki, M. Viviani, 
A. Kievsky, S. Rosati, J. Carlson, Steven C. Pieper, 
R. B. Wiringa, P. Navratil, B. R. Barrett, 
N. Barnea, W. Leidemann, and G. Orlandini, 
Phys. Rev. {\bf C64}, 044001 (2001).




\bibitem{ocm}S. Saito, Prog. Theor. Phys. {\bf 41}, 705 (1969).

\bibitem{Kukulin84}V. I. Kukulin, V. N. Pomerantsev, Kh. D. Razikov,
V. T. Voronchev, and G.G. Ryzhikh, Nucl. Phys. {\bf A586}, 151 (1995).

\bibitem{c12} J.L. Visscher and R. van Wageningen, Phys. Letters 
{\bf 34B}, 455 (1971).

\bibitem{Bodmerxx} S. Ali and A.R. Bodmer, Nucl. Phys. {\bf 80}, 
99 (1980). 

\bibitem{Horiuchi}H. Horiuchi, Prog. Theor. Phys. {\bf 51}  
1226, (1974); {\bf 53}, 447  (1975).

\bibitem{Kanada79} H. Kanada, T. Kaneko, S. Nagata, and
M. Nomoto, Prog. Thoer. Phys. {\bf 61}, 1327 (1979).

\bibitem{Hasegawa71} A. Hasegawa and S. Nagata, Prog. Theor. Phys.
{\bf 45}, 1786 (1971).

\bibitem{Yama94} Y. Yamamoto, T. Motoba, H. Himeno, K. Ikeda,
and S. Nagata, Prog. Theor. Phys. Suppl. No.117, 361 (1994).

\bibitem{Ichi99} K. Ahn {\it et al.},
AIP Conference Proceedings of the Internatinal Symposium on
Hadron and Nuclei, Seoul, 2001 
, p.180.

\bibitem{Ichi01} A. Ichikawa, Ph.D. thesis, Kyoto Univerity (2001), 
unpublished.


\bibitem{Akikawa02} H. Akikawa {\it et al.}, Phys. Rev. Lett.
{\bf 88}, 82501 (2002)

\bibitem{Motoba83} T. Motoba, H. Band\={o}, and K. Ikeda,
  Prog. Theor. Phys. {\bf 70}, 189 (1983); T. Motoba,
  H. Band\={o}, K. Ikeda, and T. Yamada, Prog. Theor. Phys.
Suppl. {\bf 81}, 42 (1985).

\bibitem{Hiyama7Li} E. Hiyama, M. Kamimura, K. Miyazaki, and
T. Motoba, Phys. Rev. {\bf C59}, 2351 (1999).

\bibitem{TamuraTanidaE419} H. Tamura, Nucl. Phys. {\bf A639},
  83c (1998); H. Tamura {\it et al.},  Phys. Rev. Letters
 {\bf 84}, 5963 (2000); K. Tanida {\it et al.}, Phys. Rev. Letters
 {\bf 86}, 1982 (2001).

\bibitem{Miyahara83} K. Miyahara, K.Ikeda, and H.  Band\={o}, Prog. 
Theor. Phys. {\bf 69}, 1717 (1983): K. Ikeda, H. Band\={o},
and T. Motoba, Prog. Theor. Phys. Suppl. {\bf 81}, 147 (1985).


\end{thebibliography}
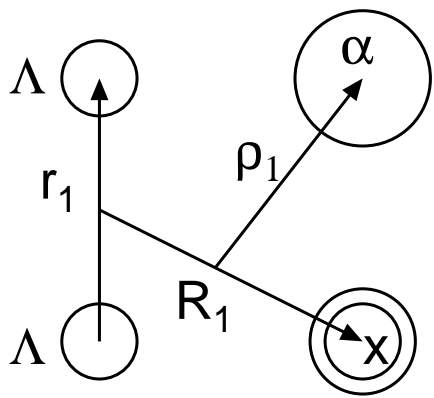
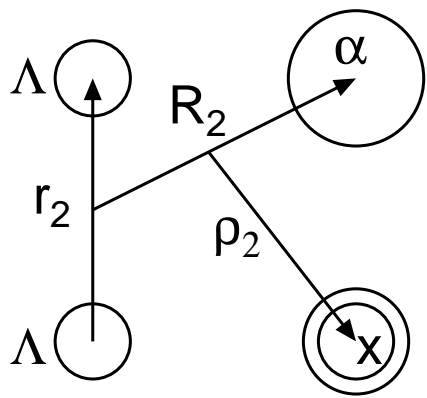
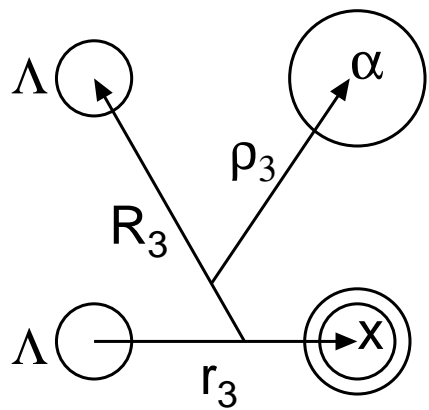
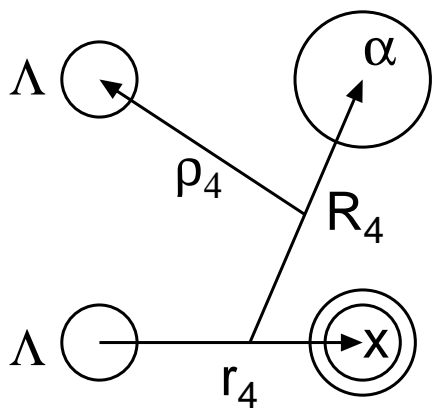
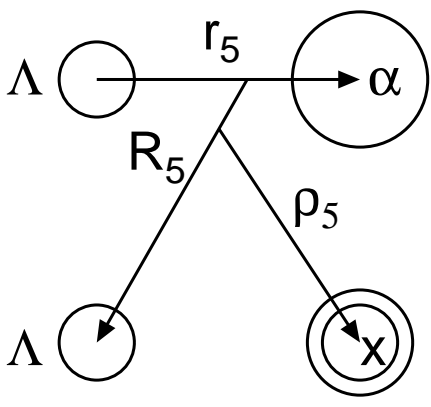
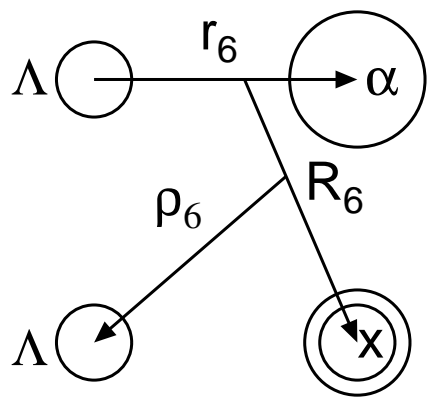
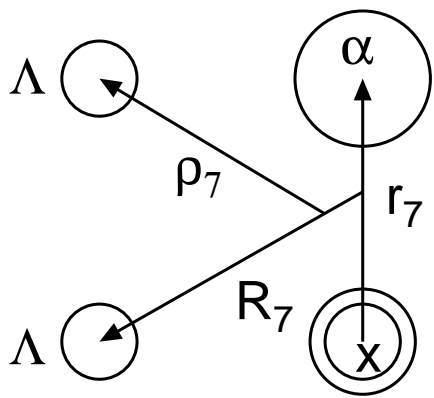
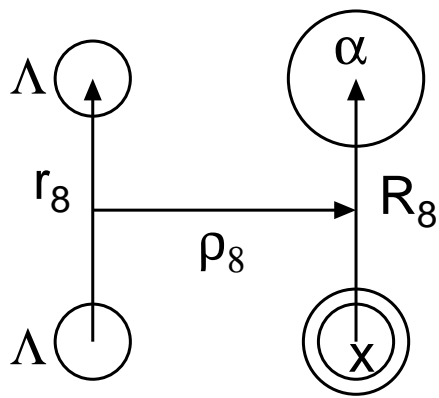
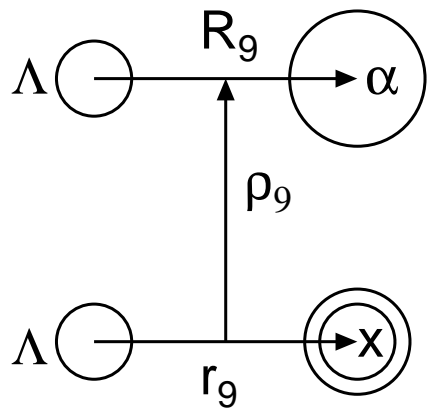
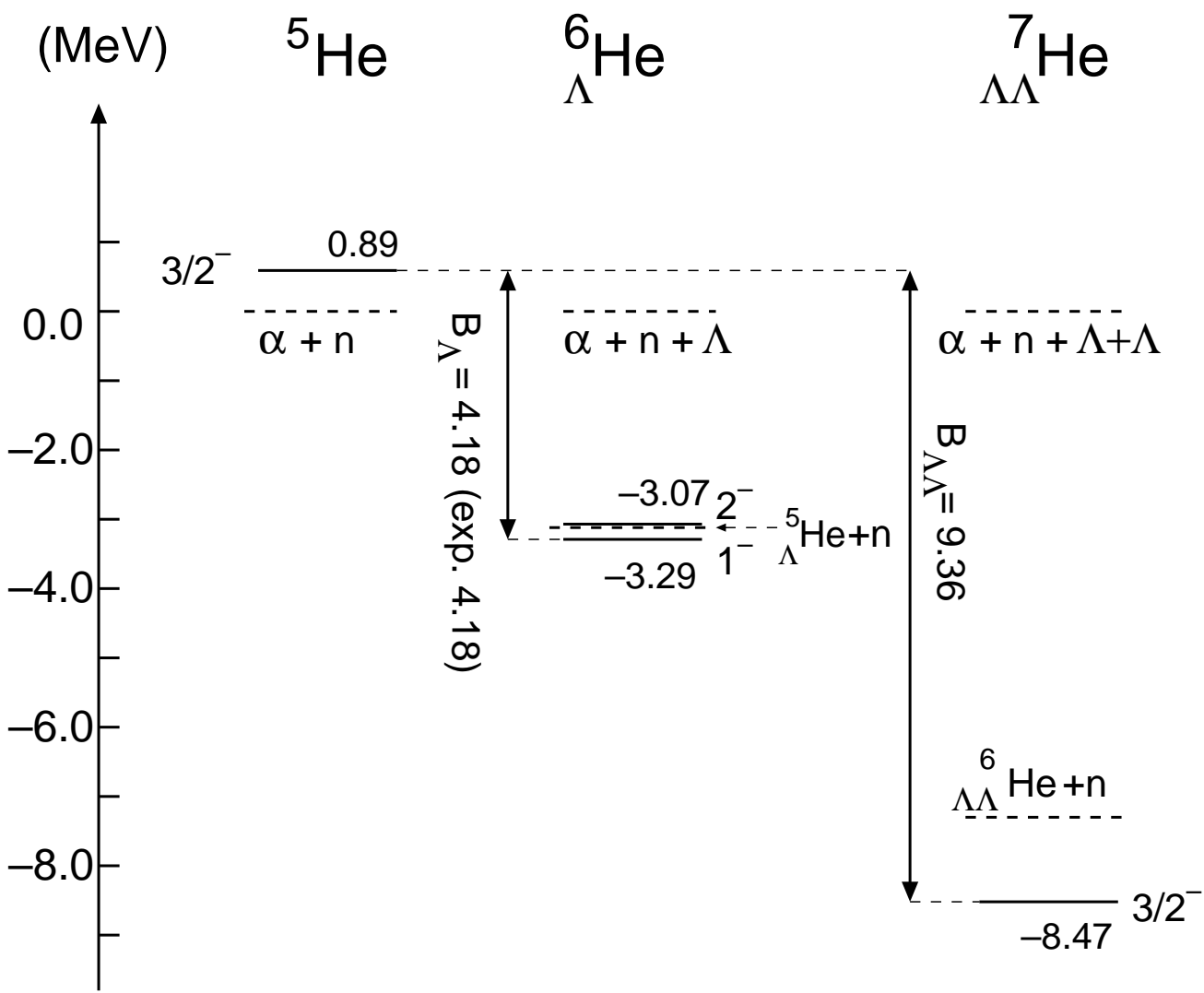
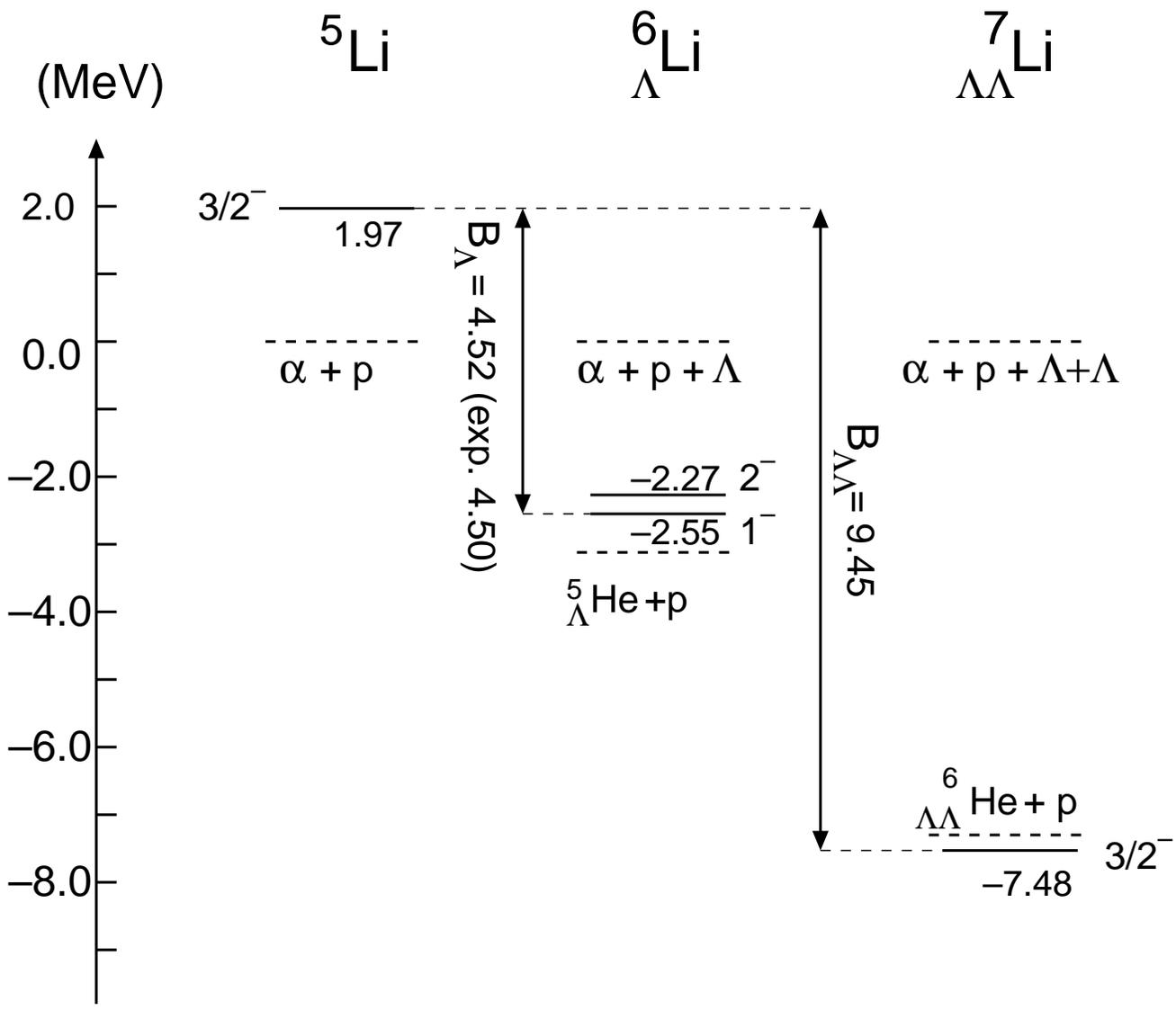
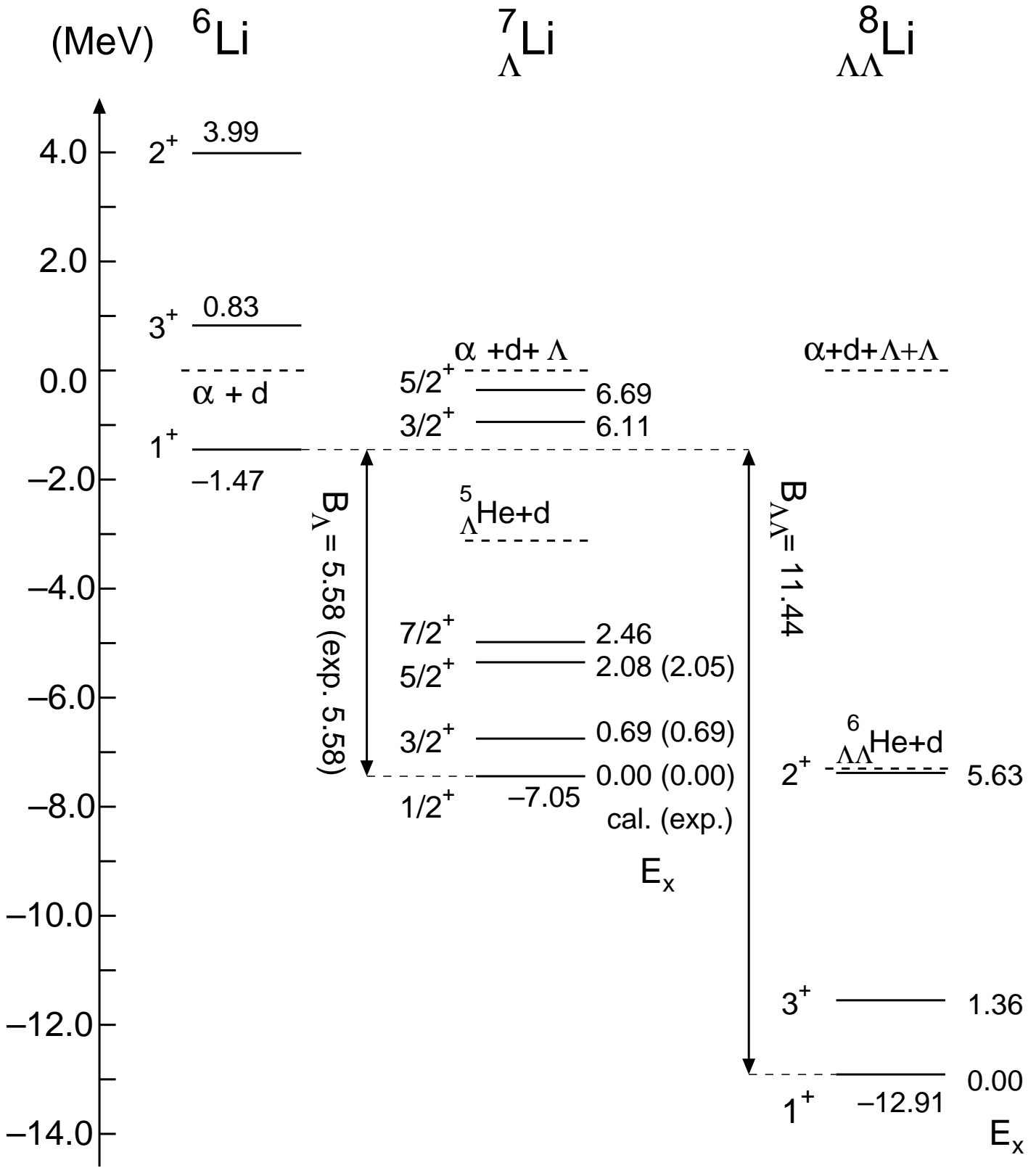
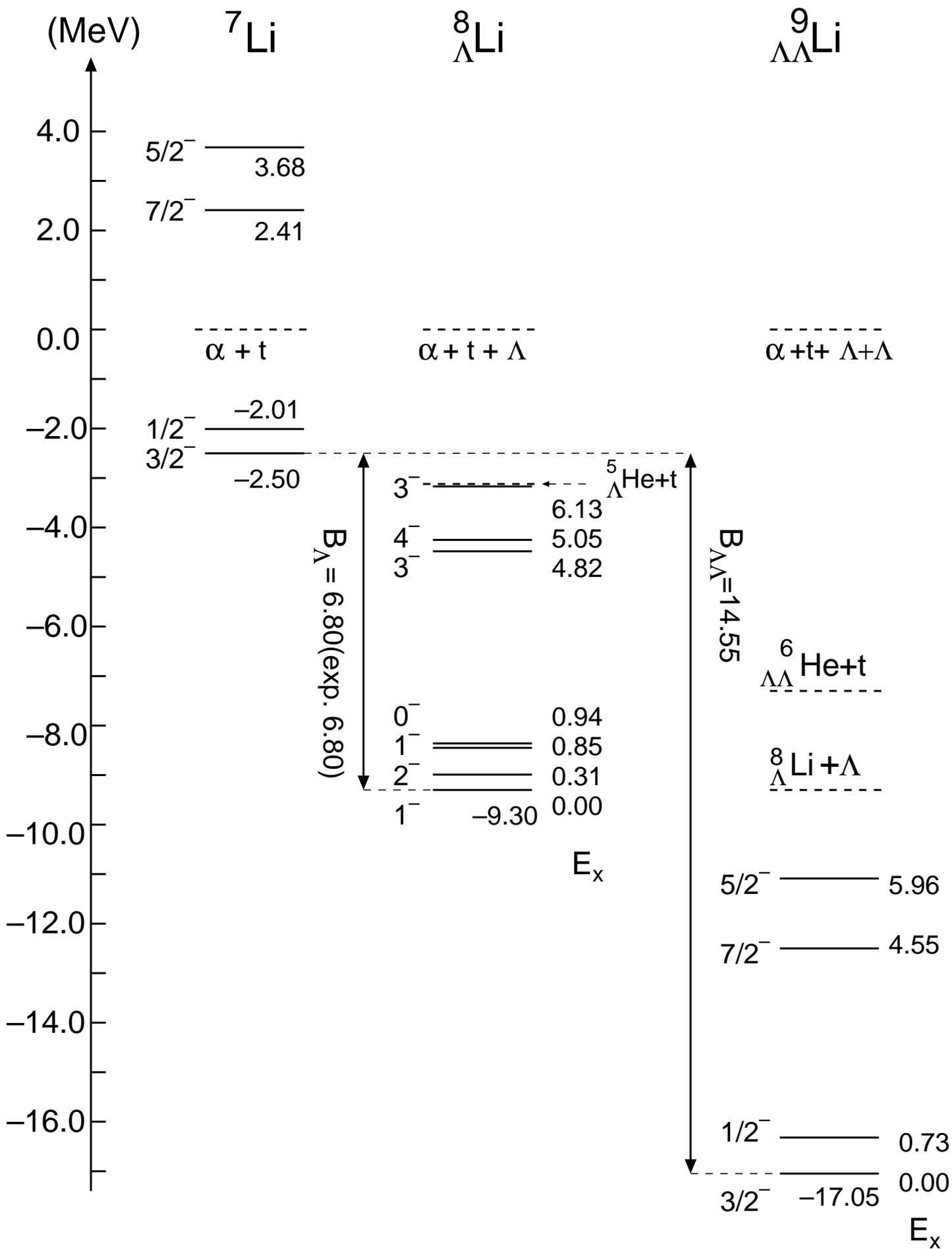
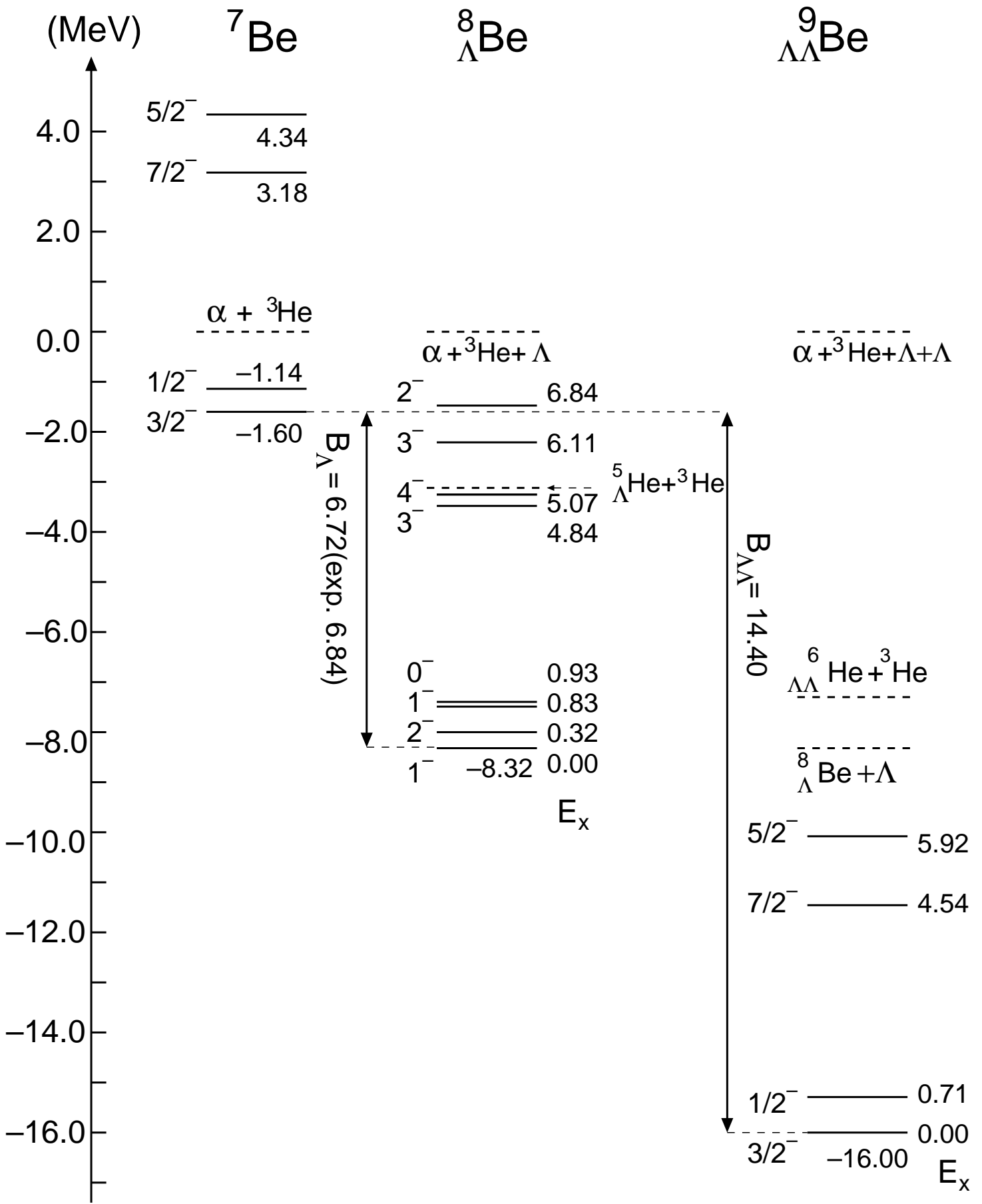
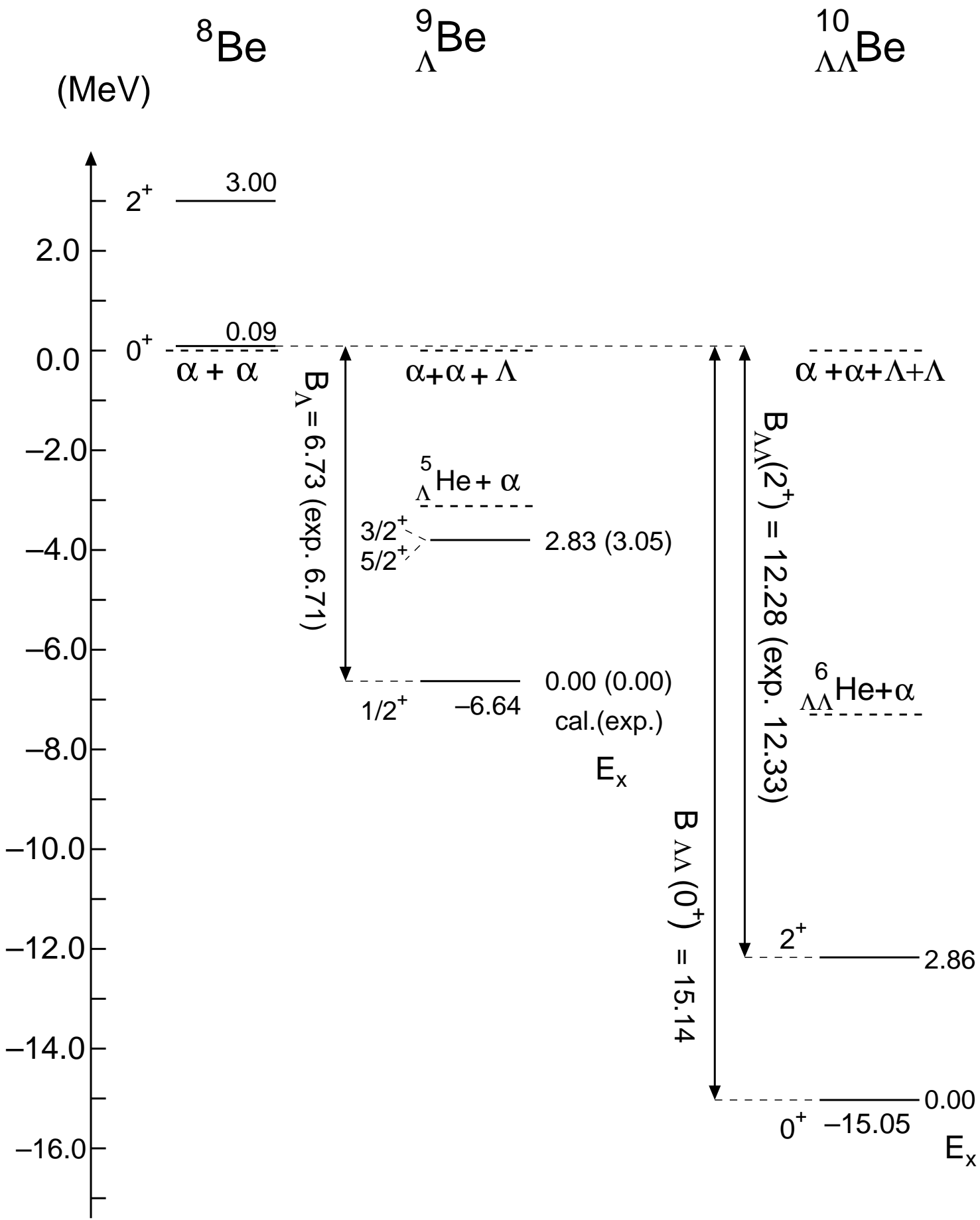
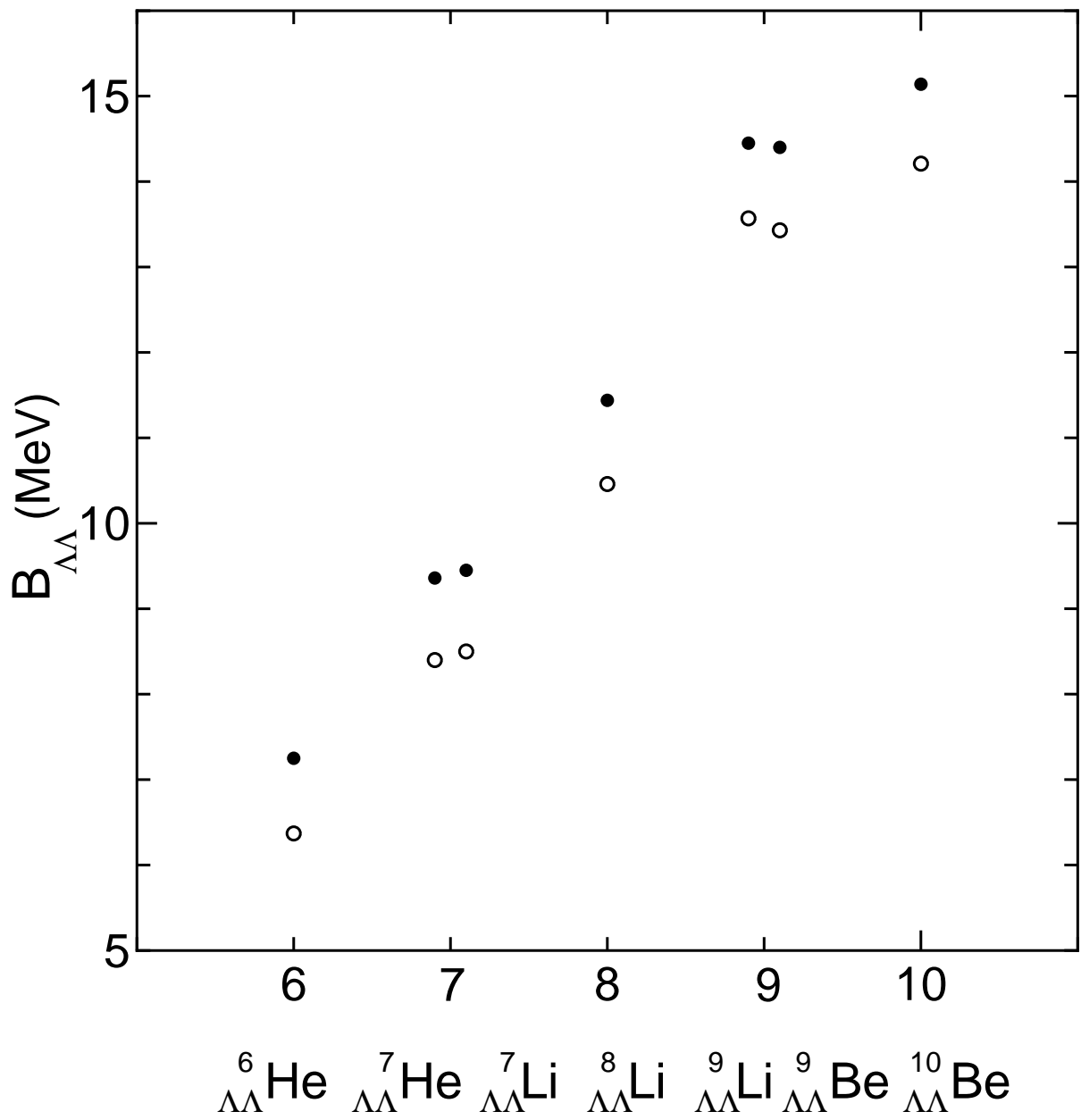
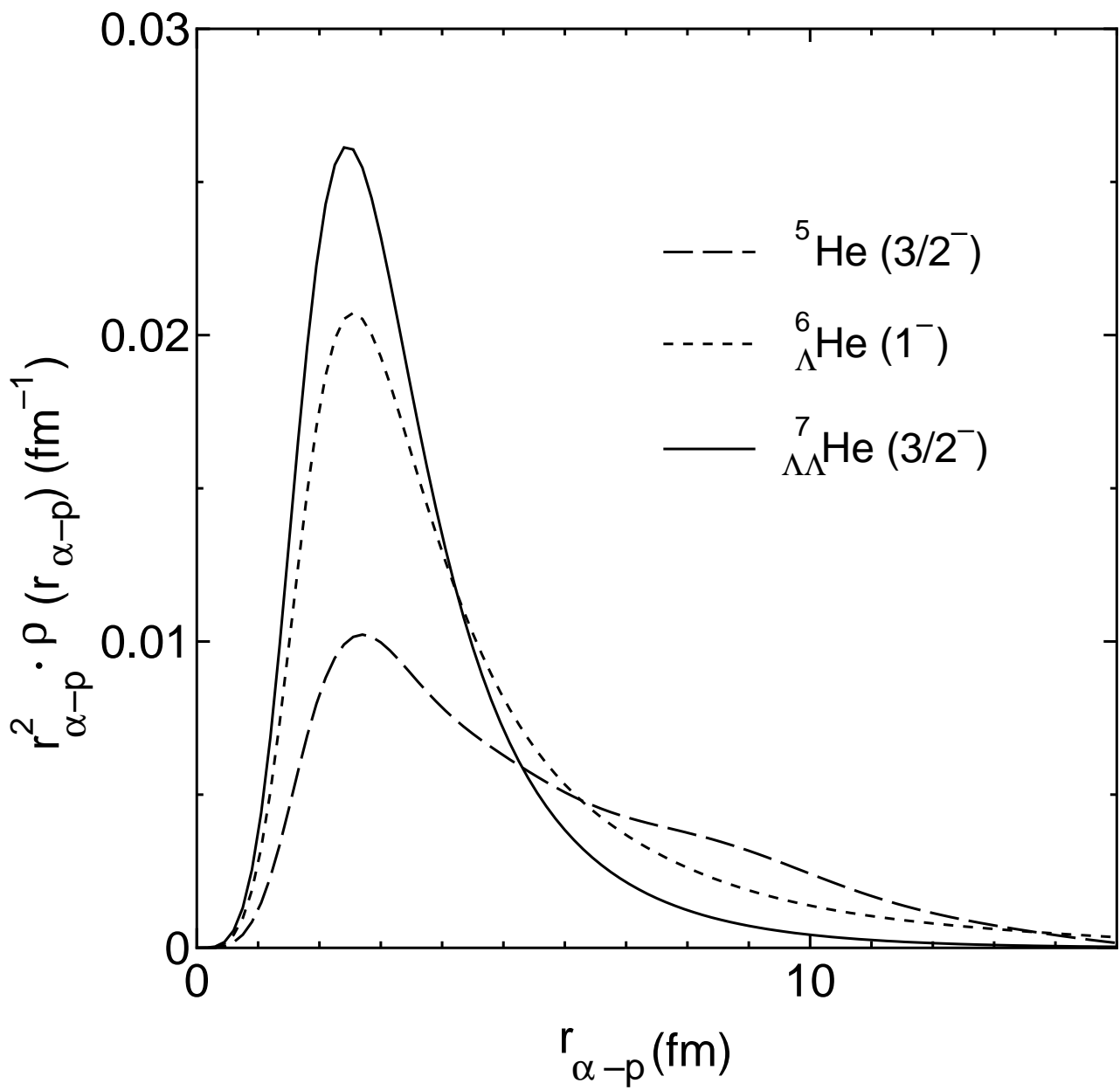
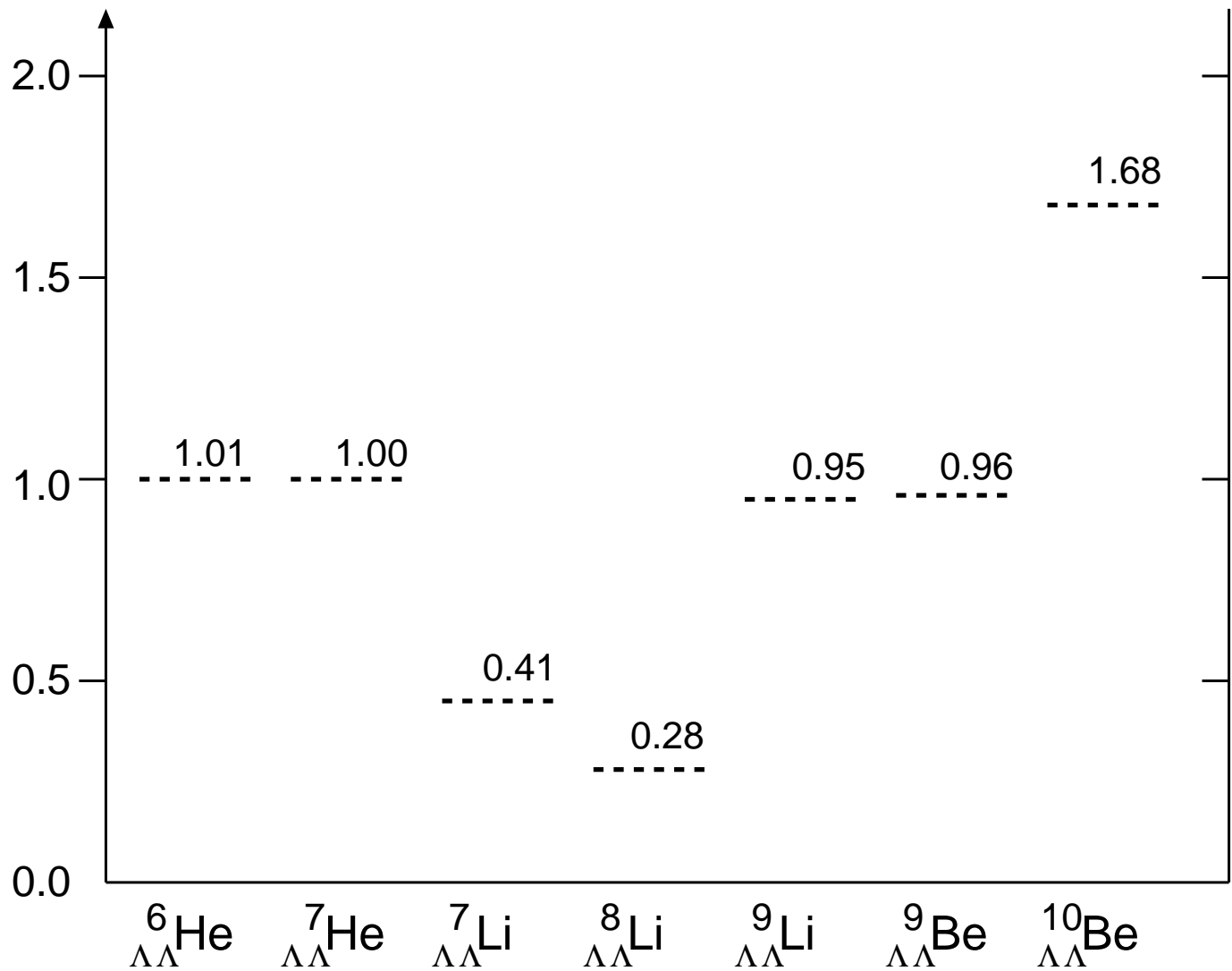
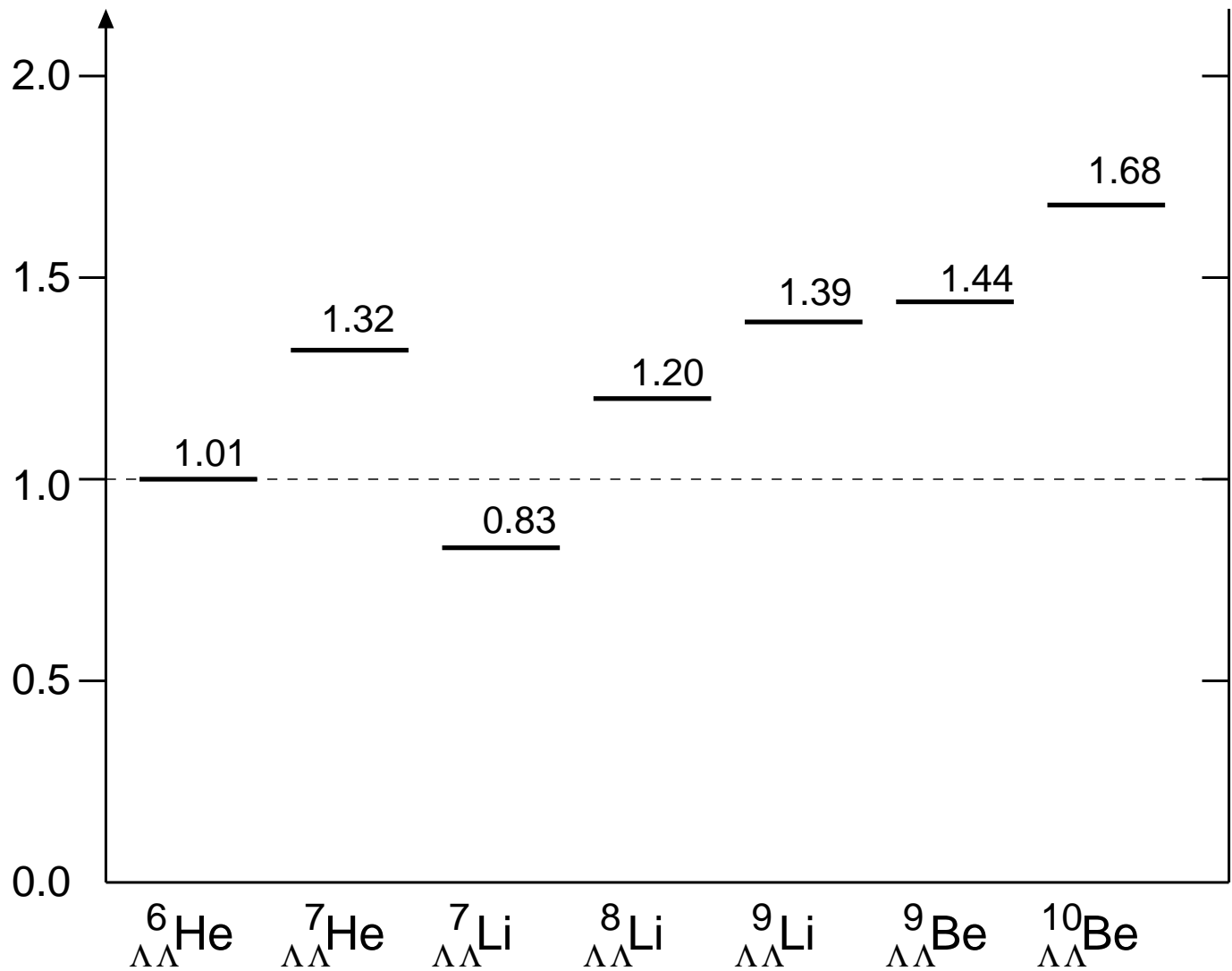
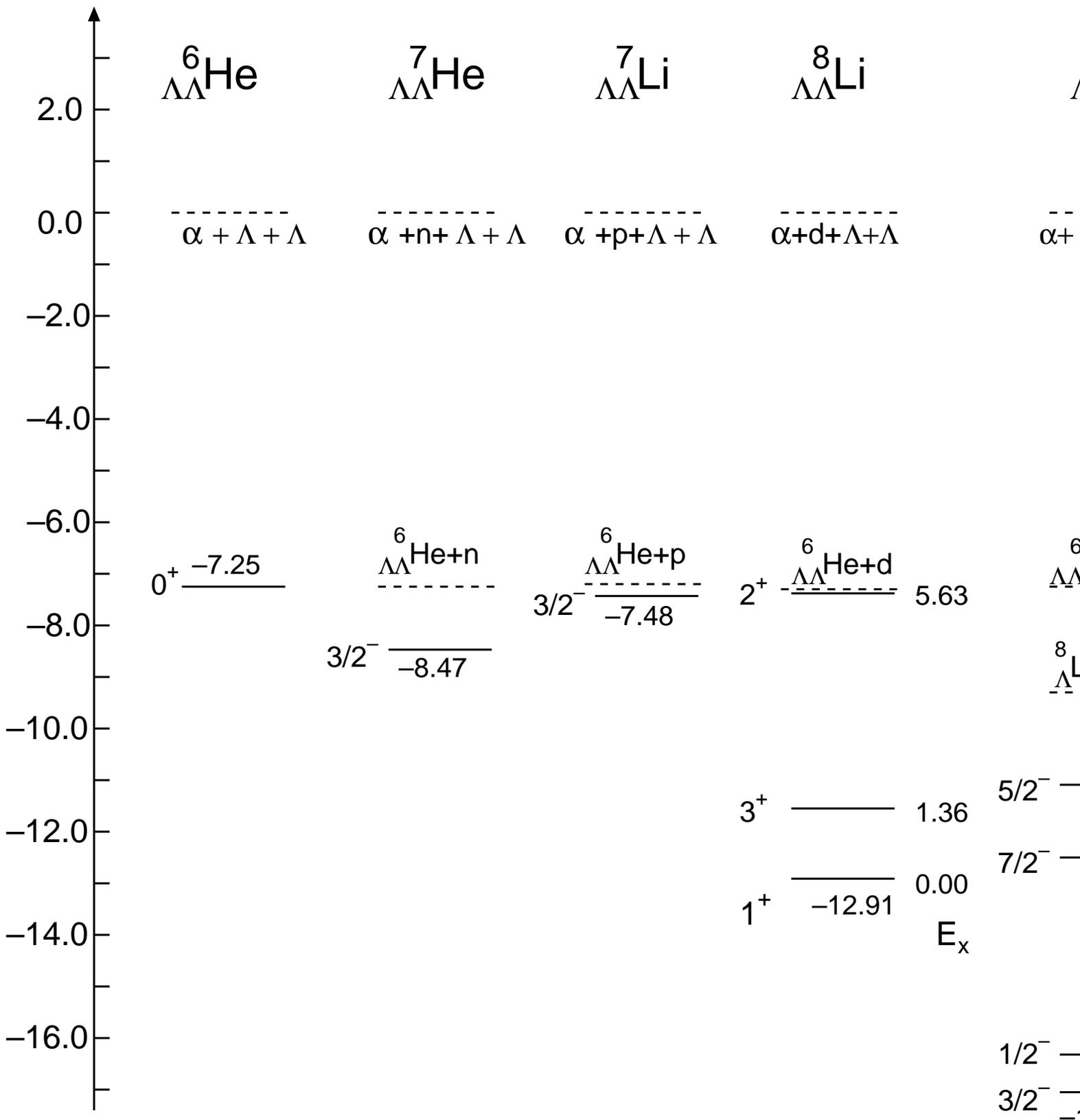
\end{document}